\documentclass[twocolumn]{30onr_art}
\usepackage{natbib}
\usepackage{ulem}
\usepackage{graphicx}
\usepackage{fixltx2e}
\usepackage{amsmath,float}
\usepackage{graphicx}
\usepackage[english]{babel}
\usepackage{epsfig}
\usepackage{color}
\usepackage{epstopdf}
\usepackage{amssymb}
\usepackage{float}
\usepackage[colorlinks=true,
            urlcolor=blue,
            linkcolor=docgreen,
            citecolor=docgreen,
            bookmarks=true,
            pdftitle={The Impact of a Deep-Water Plunging Breaker},
            pdfauthor={An Wang},
            pdfsubject={Abstract Submitted for the 30th Naval Hydrodynamics Symposium},
            pdfkeywords={Breaking Waves, Wave Impact, Air Entrainment},
	    bookmarksopen=false,
            pdfpagemode=None]{hyperref}

\definecolor{docgreen}{rgb}{0,.5,0}

\pdfoutput = 1

\begin{document}
\bibliographystyle{plain}
\title{The Impact of a Deep-Water Plunging Breaker on a Partially Submerged Cube}

\author{
An Wang$^1$, Christine M. Ikeda$^1$, David A. Drazen$^2$, \\
 Anne M. Fullerton$^2$, Thomas Fu$^2$, and James H. Duncan$^1$}

\affiliation{\small $^1$University of Maryland, College Park, Maryland, U. S. A. \\
$^2$Naval Surface Warfare Center, Carderock, Maryland, U.S.A.}

\maketitle

\section{ABSTRACT}
The impact of a plunging breaker on a partially submerged cube is explored experimentally in a wave tank equipped with a programable wave maker. The experiments are conducted with the cube (dimension $L=30.48$~cm) positioned at one streamwise location relative to the wave maker and  at three heights relative to the undisturbed water level.   A single, repeatable wave maker motion producing a breaker with a nominal wavelength of 1.18~m is used.   The water surface profile at the stream wise vertical center plane of the cube is measured with a cinematic  Laser-Induced Fluorescence technique and the impact pressures on the front face of the cube are measured with piezoelectric dynamic pressure transducers. The surface profile measurements and the impact pressure measurements are synchronized. When the cube is positioned vertically so that its bottom face is at either $0.5L$ or  $0.25L$ below the undisturbed water surface, the water surface profile behaviors are basically similar with a nearly circular arc forming between the water contact point on the front face of the cube and the wave crest.  As the impact proceeds, this arc shrinks to zero size and creates a fast-moving vertical jet in a manner similar to that found in previous studies of wave impact on bottom-mounted vertical walls.  In the case where the cube is one quarter submerged, a small jet also forms at the crest and impacts the front face of the cube just before the circular arc reaches zero size.  When the bottom of the cube is located at the undisturbed water level the wave impact is dramatically different.  In this case, it appears that a packet of air is entrapped during the impact and the surface pressure subsequently oscillates with a frequency of about 2,000 Hz.

\section{INTRODUCTION}
\begin{figure*}[!htb]
\begin{center}
\begin{tabular}{c}
\includegraphics[width=0.98\linewidth]{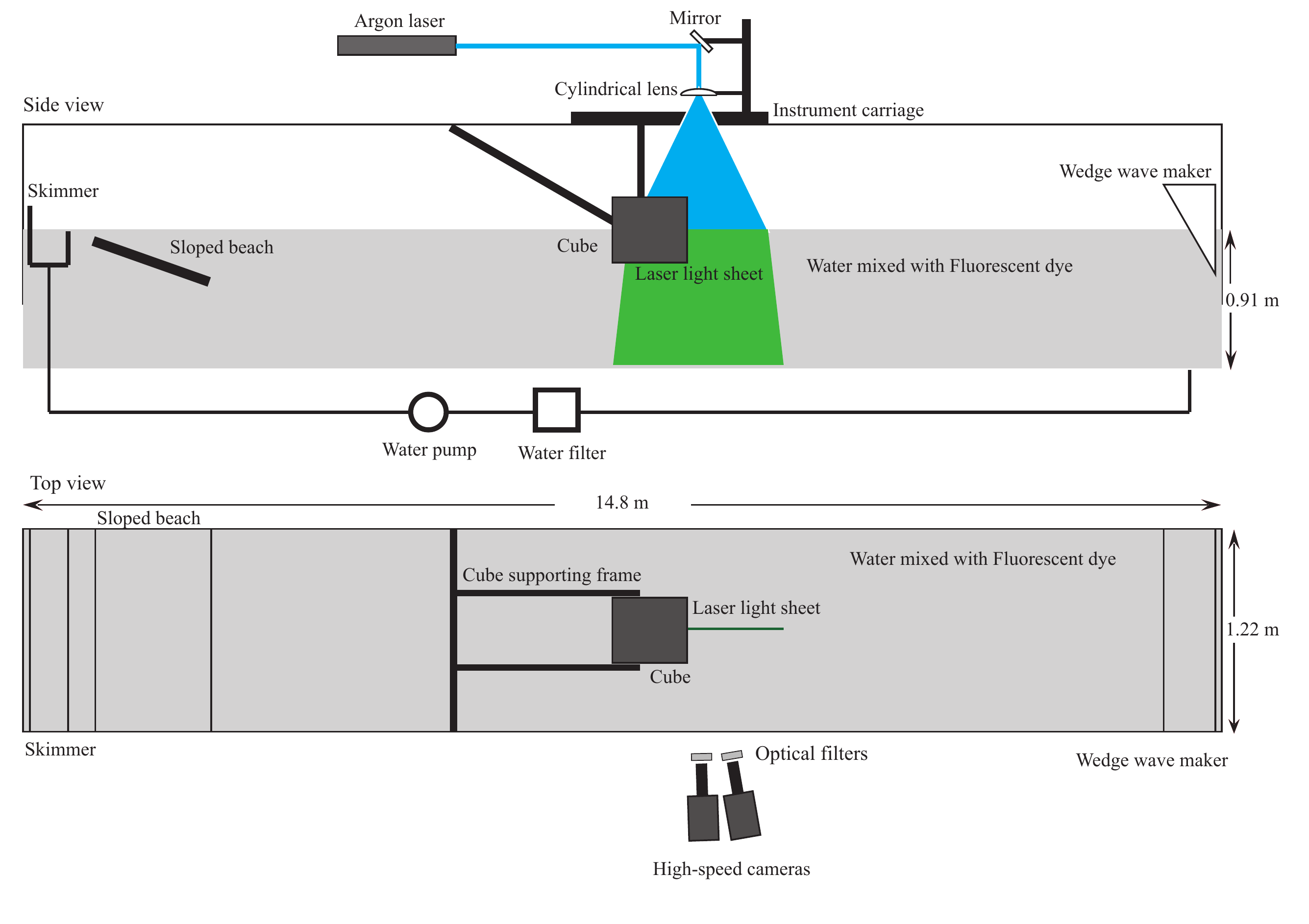}\\
\end{tabular}
\end{center}
\vspace*{-0.2in} \caption{Schematic of experiments.} \label{fig:wavetank}
\end{figure*}

The impact of breaking waves on structures is a classic problem in ocean engneering applications. Breaking wave impact is very difficult to deal with in both theory and experiments, due to its nonlinear, transient and singular nature. Plunging breakers are a class  of breaking waves that form a jet at the front face of the crest as the wave steepens. The  jet eventually overturns and plunges into the water, leading to the generation of turbulence, the ejection of droplets into the air and the entrainment of air bubbles into the water.  Plunging breakers have been studied by many researchers. For example, Longuet-Higgins and Cokelet (1976), Dold and Peregrine (1984), Dommermuth et al.\ (1988)  did numerical simulations of plunging breakers, while experiments have been performed by  Rapp and Melville  (1990) and  Ting and Kirby (1995). Chanson (2002)  and others have studied  air entrainment in plunging breakers. 

When a structure is installed near the breaking location, the breaking behavior is, of course, dramatically different  from  a plunging breaker in open water.   

The impact of a two-dimensional plunging breaker on a wall with relatively large vertical extent and width compared to  the wave height can be divided into several regimes based on the streamwise position of the wall relative to the breaking location in the absence of the wall, see for example Chan and Melville (1988) \ and  Peregrine (2003) . If the wall is located a wavelength or so downstream of the breaker location, the plunging jet from the crest of the breaker overturns and plunges back into water before the breaking crest reaches the wall.  In this case, a significant amount of air can be entrained into the  water and this can change the average compressibility of the fluid and  the pressure distribution on the wall. According to Chan and Melville (1988), double pressure peaks  are possible at this condition in cases where the secondary jet that originates from the impact of the main plunging jet into the water can impact the wall before the turbulent breaking region of the wave. If the wall is located at a position far upstream of the breaking location, the  breaker does  not fully developed before reaching the wall and the impact phenomenon is relatively weak without violent impact pressure. 
At a critical wall position in between the above two cases, the breaker is just about to form a  jet as it reaches the wall and at the same time the contact point of the water surface on the wall is moving upwards and meets the crest.  In this case, the impact does not involve breaking as the water surface profile undergoes a  phenomenon called ``Flip-through", see Peregrine (2003) , in which a high-speed vertical jet is formed and very high wall pressures are developed.

Among the features of wave impact phenomenon, the violent impact pressure has received a great deal of attention from researchers. The mechanism that generates the high impact pressure is still not well understood. Viscosity is considered an unimportant factor until after the impact because the short time scale of wave impact doesn't allow viscous effects to diffuse from the boundary layer into the bulk of the flow. Surface tension is only important during the formation of the plunging jet, which has a large curvature at its tip.  Gravity is also not relevant because at the moment of impact the acceleration of the near-surface water is many times the gravitational acceleration. Depending on the shape of the wave crest just before the  impact, the pressure magnitude is likely to  be dominated by the rapid changes in fluid inertia and the sudden jump in free surface topology.

A number of investigators have explored the wave impact pressures in laboratory and field experiments as well as in numerical and theoretical investigations.   Bagnold (1939) developed a theoretical model to estimate the impact pressure. Full-scale field measurements of impact pressure on a seawall were conducted by Blackmore and Hewson (1984) . Cooker and Peregrine (1990) predicted the 2D wave impact pressure by using a pressure impulse model and solving the boundary value problem without considering the air entrainment. Kirkg{\"o}z (1991, 1995) measured the impact pressure by breaking waves on a wall (reaching the tank bottom) with various slopes. It was found that the impact pressure on a back-sloping wall is larger than that on a vertical wall. Cooker and Peregrine  (1995) proposed a pressure-impulse theory for modeling the impact pressure and velocity field. Several impact examples were calculated with various geometries. They argue that the pressure impulse is not sensitive to the shape of the wave. Wood and Peregrine  (1998) used the pressure-impulse theory in studying three-dimensional wave impact on vertical wall. Based on their calculations, it is found that if the impact width is less than twice the water depth, 3D effects are significant.  A series of detailed experiments on 2D wave impact on a vertical wall was conducted by Chan and Melville (1988). In their work, as mentioned above, the impact characteristics are divided into several regimes based on the streamwise position of the wall and the behavior of the pressure. The variability of the impact pressures over repeated runs under the same condition is shown. 

As mentioned above, in some cases air entrainment occurs during or before the wave impact.  During the wave impact, a large air pocket is entrapped  when the plunging jet directly hits the wall; the shape and volume of this air pocket depends on the shape of the wave crest at the moment just before impact on the structure.    If the plunging jet falls back into the water before impacting the structure, the entrapped air pocket can break up into many small bubbles, forming an air-water mixture that subsequently collides with the wall. The mixture has decreased density and increased compressibility, thus decreasing the speed of sound dramatically.  In either of the above cases, the dynamics and compressibility of the air can influence the behavior of the fluid and the wall pressure.

Air entrainment during wave impact has been studied in experiments and theory.   In an early study by Bagnold  (1939), the cushion effect of the air phase is discussed. It is mentioned that the thickness of the air cushion is important and that the thickness is not uniform along the impact zone due to the irregularity of the wave crest.  This results in a variation of pressure along the impact zone. Zhang et al.\ (1996) simulated  plunging wave impact with a mixed-Eulerian-Lagrangian boundary-integral method. In the simulations, the entrapped air is taken into account by a bubble pressure modeled with a polytropic gas law.  Bullock et al.\ (2001) studied the effect of an air cushion on wave impact for both fresh water and sea water. Peregrine and Thais  (1996) developed a method for estimating the effect of an air cushion on the wave impact pressure. In their work, the aerated water is assumed to be an incompressible liquid with a void fraction, representing homogeneously distributed small bubbles in water.  Their results show that the Mach number of the incoming flow and the severity of the wave can affect the pressure reduction due to the air cushion during impact. It is also emphasized that even a small fraction of air dispersed as small bubbles in the flow can decrease the impact pressure significantly. Walkden (2001)investigated the wave impact pressure on a caisson breakwater on both the front side and back side during overtopping impact events. It was found that the air  trapped by a plunging breaker can reduce the pressure impulse during an overtopping impact event.

Air entrainment  makes the wave impact phenomenon difficult to be scaled. Traditional Froude scaling can overestimate the impact pressure in cases with air entrainment, see Bullock et al.\  (2001), because the cushion effect of the air phase is not taken into account.  In the model proposed by Zhang et al.\  (1996), see above, the air phase oscillations are modeled as a spring-mass system, which can simulate the oscillation of the air packet. 

For an object with relatively small dimensions compared to the wave length, the impact process will be more complex than that for the previous investigations of wave impact on large scale walls.   For one thing, when the bottom of the object is close to the mean free surface, it can be above the instantaneous free surface level for a short time during the impact process.  When the water surface subsequently impacts the underside of the object, upstream propagating disturbances can be generated on the water surface.  Also, as the wave approaches the object, vorticity can be generated at the edges of the object and this vorticity can further influence the impact process.  Finally, the submerged portion of  the structure can have a large influence on the wave behavior during impact.

In this paper, an experimental investigation of the impact of a plunging breaker on a partially submerged cube is presented.  The wave impact for one wave maker motion and three cube positions (three vertical positions at one horizontal distance from the wave maker) is studied.  The following presentation of the experiments is divided into three sections.  The experimental details as described in the following section.  The results are presented and discussed in the third section of the paper.    The conclusions are presented in final section.

\section{EXPERIMENTAL DETAILS}

\begin{figure*}[!htb]
\begin{center}
\begin{tabular}{c}
\includegraphics[width=0.72\linewidth]{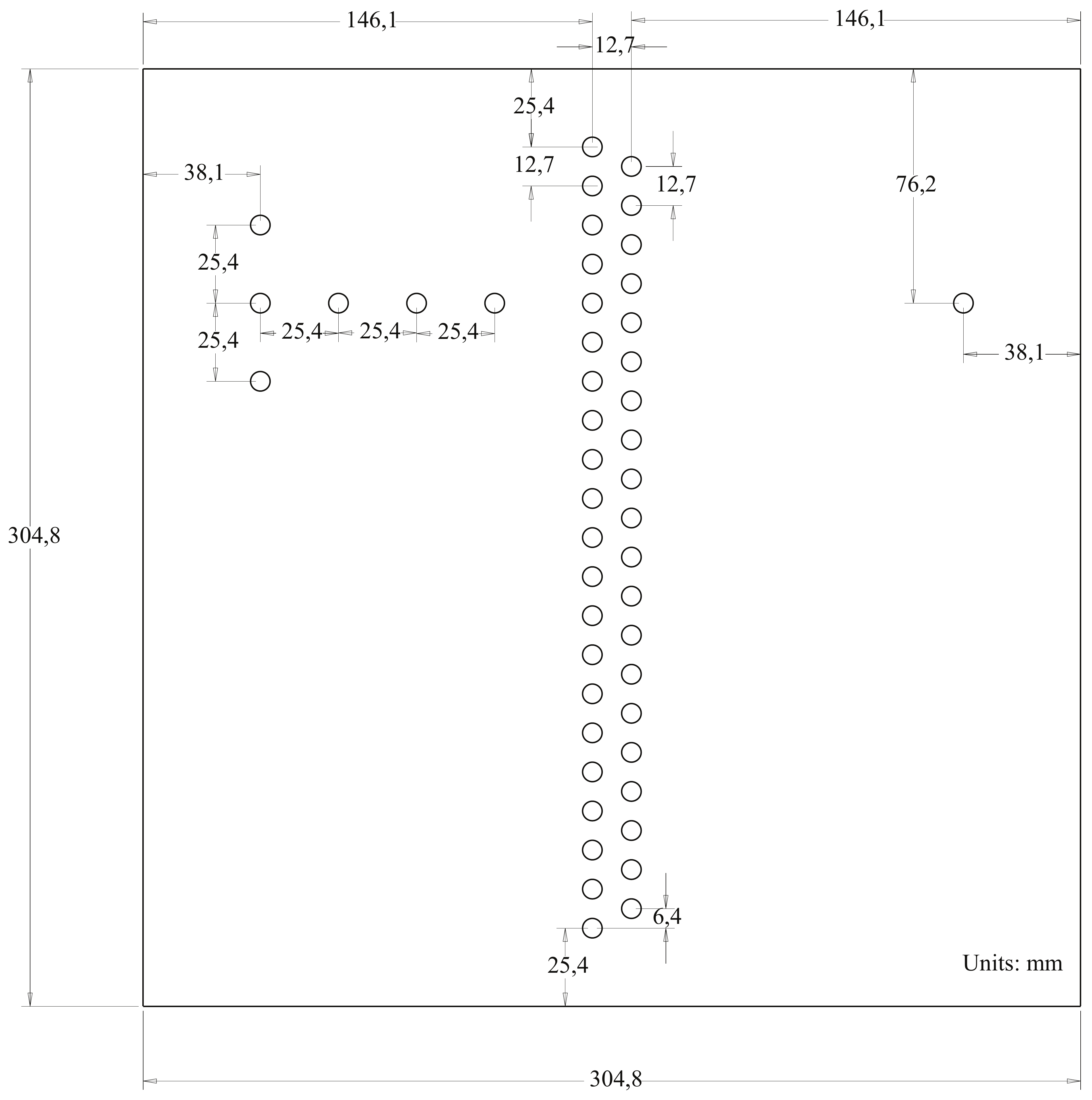}\\
\end{tabular}
\end{center}
\vspace*{-0.2in} \caption{Layout of mounting holes for pressure transducers at the front face of the cube. Each circle represents a mounting hole. } \label{fig:cubefrontface}
\end{figure*} 
The experiments are conducted in a wave tank that is 14.8 m long, 1.22 m wide and  2.2 m tall with a water depth of  0.91~m, see Figure~\ref{fig:wavetank}. A 30.48-cm cube made of six  Aluminum plates is placed in the tank as shown in the figure.  The cube is rigidly supported by an aluminum frame that is mounted to the top of the tank.

The water surface profile in a vertical  plane at the center of the cube  is measured with a cinematic laser-induced-fluoresence technique. The illumination for these movies is provided by a downward-directed argon ion laser beam, which  is focused to a point just above the mean water surface by two spherical lenses and expanded in the long center plane of the tank by a cylindrical lens. During the experiments, the water is mixed with fluorescein dye.  The dye illuminated by the light sheet fluoresces and two cameras record the intersection of the light sheet and the water surface from the side as shown in Figure~\ref{fig:wavetank}.  Long-wavelength-pass optical filters are placed in front the camera lenses so that only the light from the fluorescing dye reaches the camera sensors.  In this way, the cameras record the profile of the water surface in the center plane of the cube during the wave impact.   One of the high-speed cameras is set to record 2560 pixel $\times$1600 pixel images at  a rate of 1500 fps and a field of view of about 62~cm, while the other camera is records 1280 pixel $\times$ 800 pixel images at a rate of 4500 fps and a field of view of about 14~cm.  The air-water interface  in the images is determined  by digital edge-detection techniques. 

The plunging breaker is generated by a programable wave maker located at one end of the tank. The wave maker consists of a wedge which oscillates in the vertical direction. The wedge is driven by a ball-screw and servo-motor system which is in turn controlled by a computer-based PID controller with position and velocity feedback. Repeated runs of the wave maker with the same input motion result in a 0.1\% root-mean-square error in wedge position at the time of the peak displacement from the mean water. 

The plunging breaker is generated by using the dispersive focusing technique described in Longuet-Higgins (1974), Rapp and Melville (1990) and Duncan et al (1999). The wave maker generates a wave packet with wave components at different frequencies (average frequency of the wave packet is $f_{0}=1.15$ Hz). According to the dispersion relation from the deep water linear wave theory, wave components with different frequencies travel at different phase velocities. Therefore, the frequencies can be controlled in a way such that all the components are focused at a desired streamwise location, where a plunging breaker forms when the amplitudes of the initial waves are large enough. The wave maker motion is described in the following equation.
\begin{equation}
z_{w} = w\left(t\right)\frac{2\pi}{N}A\sum_{i=1}^{N}\frac{1}{k_{i}}\cos\left[x_{b}\left(\frac{\omega_{i}}{\bar{c_{g}}}-k_{i}\right)-\omega_{i}t+\phi\right]
\end{equation}
where $N$ is the number of wave components and $N=$ 32, $A$ is a constant representing the amplitude of the wavemaker motion, $x_{b}$ is the streamwise position of the breaking event measured from the back of the wavemaker, $k_{i}$, $\omega_{i}$ are the wavenumber and frequency of the $i^{th}$ component. By linear theory, $k_{i}=\omega_{i}^{2}/g$. $\bar{c_{g}}$ is the average group velocity of $N$ components and $\phi$ is the phase shift, which is chosen to be $\pi/2$. The frequencies are equally spaced, $\omega_{i}=\bar{\omega}-\Delta\omega/2+(i-1)\Delta\omega/(N-1)$, where $\bar{\omega}$ and $\Delta\omega$ are constants, $\bar{\omega}=2\pi f_{0}$ and $\Delta\omega=0.77\bar{\omega}$. The window function $w (t)$ is introduced in order to give the wavemaker zero motion at times when the summation of all the components generates a very small motion:
\begin{eqnarray}
w\left(t\right) &=& 0.25\left(\tanh\left(\beta f_{0}\left(t-t_{1}\right)\right)+1\right)\times \nonumber\\
& &\left(1-\tanh\left(\beta f_{0}\left(t-t_{2}\right)\right)\right)
\end{eqnarray}
The value of the window function is zero when $t<t_{1}$ or $t>t_{2}$ and close to 1 when $t_{1}<t<t_{2}$. $t_{1}$ and $t_{2}$ are defined as
\begin{equation}
t_{1}=x_{b}\left(1/\bar{c_{g}}-1/c_{N}\right)
\end{equation}
\begin{equation}
t_{2}=x_{b}\left(1/\bar{c_{g}}-1/c_{1}\right)
\end{equation}
 so that the wavemaker motion can allow the fastest component (with smallest frequency) and the slowest component (with largest frequency) to travel to the desired breaking location. $\beta$ is a constant that determines the rise rate of the window function, chosen as 5.0.

All of the experiments described herein were performed with a single wave maker motion that results in a plunging breaker without the cube in the wave tank.  The characteristic wave length of this breaker can be calculated by $\lambda_{0}=2\pi g/\bar{\omega}^2=1.181$ m. $x_{b}=7.15\lambda_{0}$,  $\Delta\omega/\bar{\omega}=0.77$. By linear theory, the group velocity of each wave component is $c_{gi}=0.5\omega_{i}/k_{i}$. The average group velocity is $\bar{c_{g}}=\frac{1}{N}\sum_{i=1}^{i=N}0.5\omega_{i}/k_{i}=$ 0.7183 m/s. The average phase velocity is $\bar{c_{p}}=2\bar{c_{g}}=$ 1.4366 m/s.

The impact pressure is measured by an array of 24 piezoelectric pressure transducers manufactured by PCB Inc. These sensors have a resolution 0.007~kPa, a rise time of less than 1.0~$\mu$s, a resonant frequency above ~500 kHz and a temperature sensitivity coefficient less than 0.054$\%/^\circ$C. The diameter of the measurement face of the transducers is 0.554 cm.  The front face of the cube was machined with 48 threaded mounting holes for these transducers, see Figure~\ref{fig:cubefrontface}.  The unused 24 mounting holes are filled with dummy transducers. The mounting holes are machined so that there is a 0.508 mm recess between the measurement surface of the transducers and the front face of the cube. The recessed space is filled with insulating grease in order to delay thermal effects on the sensor readings because of changes in sensor temperature when initially dry sensors become wet during wave impact.

The positions of the mounting holes for the transducers are shown in Figure \ref{fig:cubefrontface}. Two columns of mounting holes are located symmetrically about the vertical center line of the cube front face. The mounting holes on these two columns are staggered, similar to the experiments by Chan and Melville (1998), such that the smallest vertical distance between two pressure transducers is 0.635 cm. The horizontal distance between the two columns is 1.27 cm. There are some mounting holes distributed along the horizontal direction in order to verify the possible 3D effects (not discussed in this paper).

The signals from the transducers are sampled by a 14-bit Analogue-to-Digital data acquisition system at sample rates of 20~kHz in the early experiments and 900~kHz in later experiments. The measurement range of the data acquisition system is set to $\pm$0.125~V.

Experiments were performed with the above-described wave maker motion and three cube positions.  The front face of the cube was the same, 6.42~m from the back face of the wedge of the wave maker, for all three conditions while the height of the cube was varied.  The three vertical positions chosen were such that the bottom face of the cube was located at the mean water level, $0.25L$ below the mean water level and $0.5L$ below the mean water level, where $L = 30.48$~cm, the length of the one edge of the cube.  In the following, these conditions are designated as $y_b = 0$, $y_b = -0.25L$ and $y_b= -0.5L$, respectively.  At each condition, measurements were performed for 4 or more wave impacts.

\section{RESULTS AND DISCUSSION}

\begin{figure*}[!htb]
\begin{center}
\begin{tabular}{cc}
(a) & (b) \\
\includegraphics[width=0.45\linewidth]{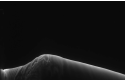} & \includegraphics[width=0.45\linewidth]{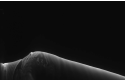} \\
(c) & (d) \\
\includegraphics[width=0.45\linewidth]{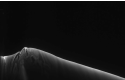} & \includegraphics[width=0.45\linewidth]{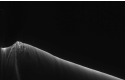} \\
(e) & (f) \\
\includegraphics[width=0.45\linewidth]{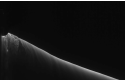} & 
\includegraphics[width=0.45\linewidth]{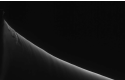} \\
\end{tabular}
\end{center}
\vspace*{-0.2in} \caption{Water surface evolution for condition $y_{b}$ = -0.5$L$ (15.24 cm submerged). The field of view of the images is 14.8 cm by 9.5 cm. (a) $t$ = -21.71 ms ($tf_{0}$ = -0.0250). (b) $t$ = -16.38 ms ($tf_{0}$ = -0.0188). (c) $t$ = -11.05 ms ($tf_{0}$ = -0.0127). (d) $t$ = -5.71 ms ($tf_{0}$ = -0.000657). (e) $t$ = -0.38 ms ($tf_{0}$ = -0.000256). (f) $t$ = 4.95 ms ($tf_{0}$ = 0.00570). The time is measured from the moment of impact, which is taken as the time when the pressure, measured by the sensor that records the highest value, reaches this highest value.  This time corresponds roughly the the time when the small circular arc in the water surface adjacent to the cube in image (f) disappears.} \label{fig:evolutions_sub15.24}
\end{figure*} 

\begin{figure*}[!htb]
\begin{center}
\begin{tabular}{cc}
(a) & (b) \\
\includegraphics[width=0.45\linewidth]{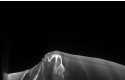} & \includegraphics[width=0.45\linewidth]{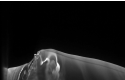} \\
(c) & (d) \\
\includegraphics[width=0.45\linewidth]{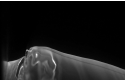} & \includegraphics[width=0.45\linewidth]{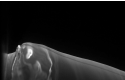} \\
(e) & (f) \\
\includegraphics[width=0.45\linewidth]{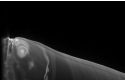} & 
\includegraphics[width=0.45\linewidth]{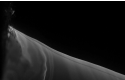} \\
\end{tabular}
\end{center}
\vspace*{-0.2in} \caption{Water surface evolution of condition $y_{b}$ = -0.25$L$ (7.62 cm submerged). The field of view of the images is 14.2 cm by 9.1 cm. (a) $t$ = -21.58 ms ($tf_{0}$ = -0.0248). (b) $t$ = -16.25 ms ($tf_{0}$ = -0.0187). (c) $t$ = -10.92 ms ($tf_{0}$ = -0.0126). (d) $t$ = -5.59 ms ($tf_{0}$ = -0.00643). (e) $t$ = -0.25 ms ($tf_{0}$ = -0.000288). (f) $t$ = 4.86 ms ($tf_{0}$ = 0.00559). The time is measured from the moment of impact.} \label{fig:evolutions_sub7.62}
\end{figure*}

\begin{figure*}[!htb]
\begin{center}
\begin{tabular}{cc}
(a) & (b) \\
\includegraphics[width=0.45\linewidth]{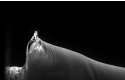} & \includegraphics[width=0.45\linewidth]{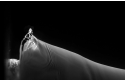} \\
(c) & (d) \\
\includegraphics[width=0.45\linewidth]{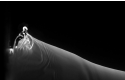} & \includegraphics[width=0.45\linewidth]{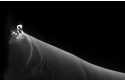} \\
(e) & (f) \\
\includegraphics[width=0.45\linewidth]{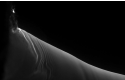} & 
\includegraphics[width=0.45\linewidth]{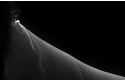} \\
\end{tabular}
\end{center}
\vspace*{-0.2in} \caption{Water surface evolution of condition $y_{b}$ = 0 (0 cm submerged). The field of view of the images is 14.2 cm by 9.1 cm. (a) $t$ = -21.66 ms ($tf_{0}$ = -0.0249). (b) $t$ = -16.33 ms ($tf_{0}$ = -0.0188). (c) $t$ = -10.99 ms ($tf_{0}$ = -0.0126). (d) $t$ = -5.66 ms ($tf_{0}$ = -0.00651). (e) $t$ = -0.11 ms ($tf_{0}$ = -0.000127). (f) $t$ = 5.00 ms ($tf_{0}$ = -0.00575). The time is measured from the moment of impact.} \label{fig:evolutions_sub0}
\end{figure*} 

\begin{figure*}[!htb]
\begin{center}
\begin{tabular}{c}
(a) \\
\includegraphics[width=0.88\linewidth]{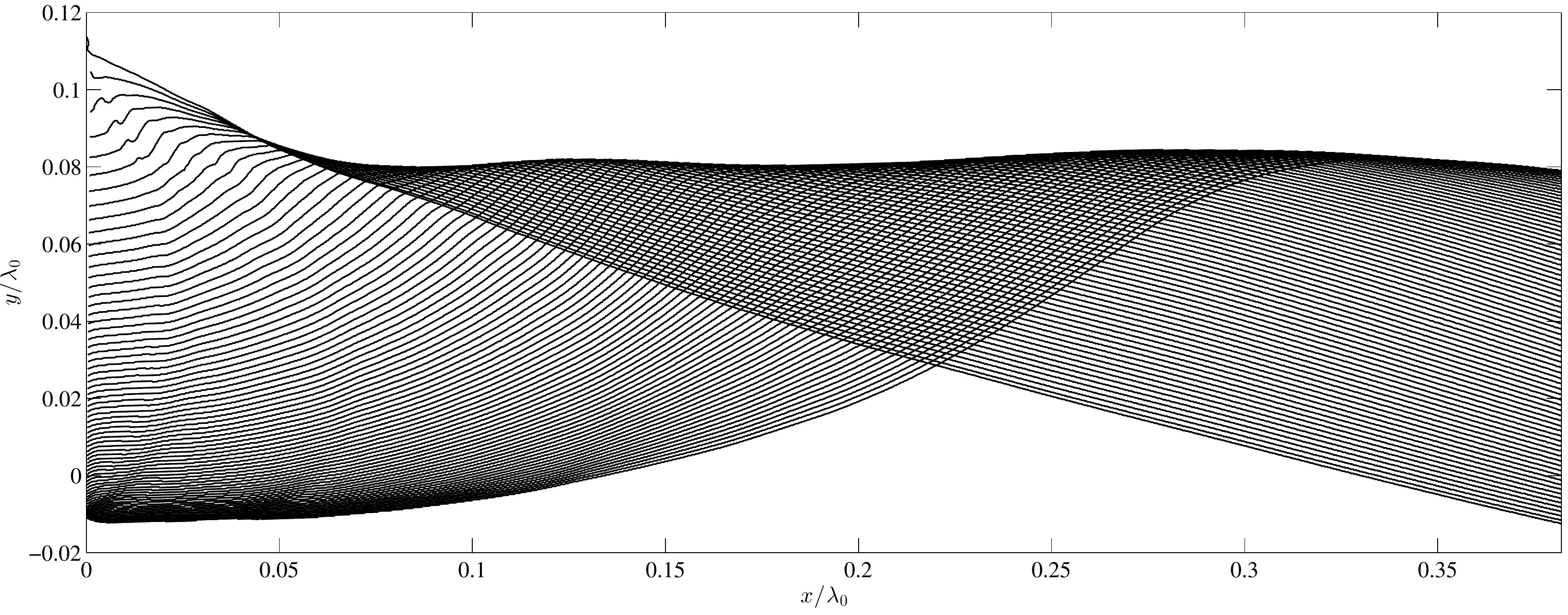}\\
(b) \\
\includegraphics[width=0.88\linewidth]{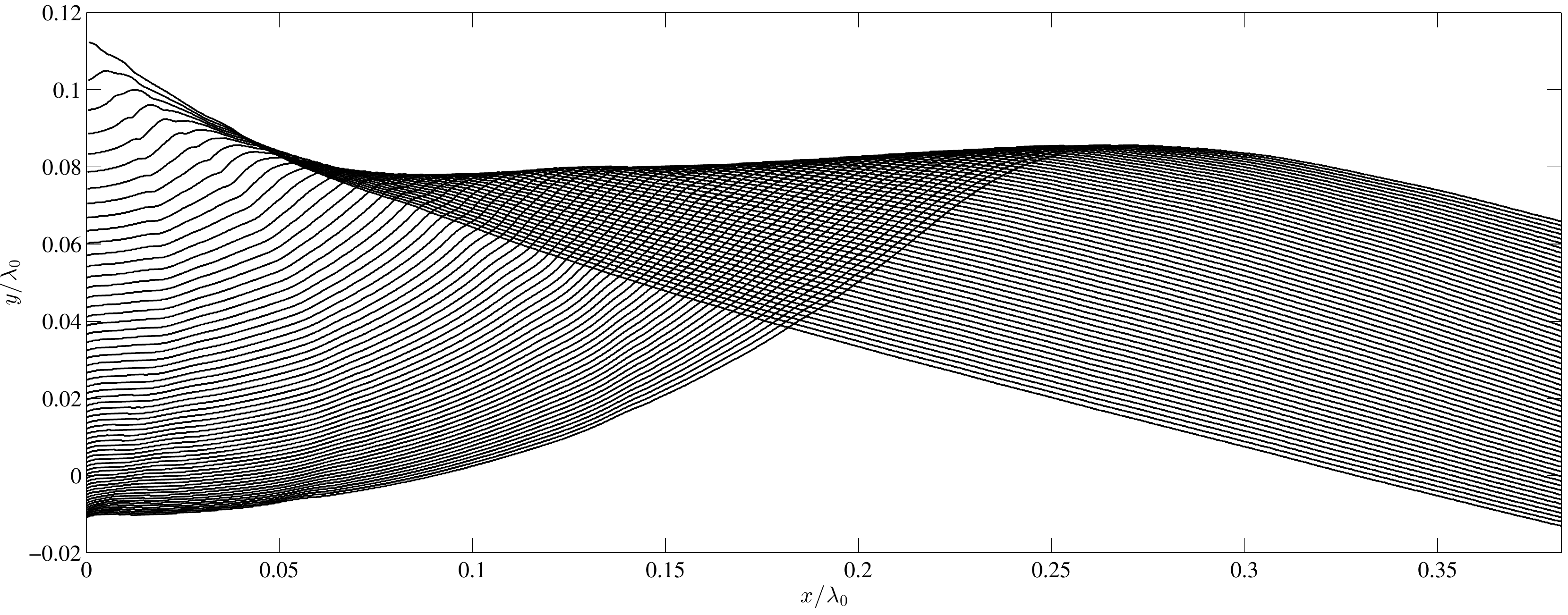}
\end{tabular}
\end{center}
\vspace*{-0.2in} \caption{Water surface profiles along the center plane of the front face of the cube (located at $x=0$ in the plots).   The profiles are captured at 1,500 frames per second and every 5th profile is plotted.  The vertical coordinate ($y$) is positive in the upward direction and $y=0$ is the mean water level. (a) Surface profiles for $y_{b}$ = -0.25$L$ (the bottom of the cube is 7.62~cm below the mean water level). (b) Surface profiles for  $y_{b}$ = -0.5$L$ (the bottom of the cube is 15.24~cm below the mean water level)} \label{fig:profiles}
\end{figure*} 

In the following, we first discuss the wave impact behavior qualitatively based on images from the high-speed movies, then examine some quantitative features of the surface profile histories and finally examine the pressure distributions on the front face of the cube.

Figure \ref{fig:evolutions_sub15.24} shows a sequence of six images from a movie of the wave impact  for $y_{b}$ = -0.5$L$ (the bottom of the cube located 15.24~cm below the mean water level). The movie was taken at  a rate of 4500 fps. As the wave crest approaches the cube, the trough ahead of the crest becomes a nearly circular arc with upward curvature extending from the contact point on the front face of the cube to  a point just ahead of the crest.  The radius of this arc shrinks to zero just after image (e) and a vertical jet is formed, see image (f).  This jet begins to slow down immediately after it is formed.  The time when the circular arc approaches zero radius is roughly defined as the moment of impact, though a more detailed definition of this time based on the pressure readings will be discussed below.  The are a number of ripples in the circular arc.  Observations from the entire movie, indicate that these ripples appear to be generated by oscillations of the flow at the  contact point earlier in the evolution of the impact as non breaking wave components pass by the cube and by the flow field at the wave crest, perhaps some kind of incipient breaking phenomenon.  

Figure \ref{fig:evolutions_sub7.62} shows a sequence of six images from a movie of the wave impact  for $y_{b}$ = -0.25$L$. The free surface behavior is similar to that in the  $y_{b}$ = -0.5$L$  case; however, the small  features on the arc between the contact point and the wave crest are more pronounced as the moment of impact is approached.    
At $t$ = -16.25 ms ($tf_{0}$ = -0.0187), a small jet develops near the wave crest.  
At the same time, the contact point is also moving upwards. The jet meets the rising contact point at nearly the same point on the front face of the cube in image (e), which is taken at roughly the moment of impact. In this case, no obvious air entrainment is observed. After the moment of impact, a high-speed vertical jet is formed and this vertical jet starts to slow down afterwards. 

Figure \ref{fig:evolutions_sub0} shows a sequence of six images from a movie of the wave impact  for $y_{b}$ = 0. The front side of the crest forms a jet which develops into a thin  sheet with relatively large size. Much of the intersection of the light sheet and the water surface between the crest and the contact point is obscured by the three-dimensional, concave shape of the jet sheet.   The contact point itself is visible and is moving upwards with high speed.  It reaches the same height as the water sheet slightly earlier than the water sheet reaches the front face of the cube. Therefore, the water sheet with horizontal momentum impacts on the high speed vertical jet. An air pocket is probably trapped between the water sheet and the high-speed vertical jet. If this air is trapped during the impact, it will likely change the bulk compressibility of the flow and influence the pressure on the front face of the cube.

Figure \ref{fig:profiles} (a) and (b) contain plots of the free surface profiles extracted from the high-speed LIF movies for $y_b = -0.25 L$ and $y_b=-0.5L$, respectively. The surface profiles for the $y_b =0$ case are not shown since the entire profile is not visible due to the three-dimensional shape of the jet, as mentioned above.  In the plots,  the origin of the coordinate system is at the intersection of the mean water level and the vertical plane of the front face of the cube. The positive $x$ direction is horizontal and opposite to the direction of wave propagation. The positive $y$ direction is upwards.

As the wave crest approaches the cube, the profiles form a nearly circular arc from the contact point to a point just downstream of the crest, as mentioned above.   The center of this circle seems to move along the $y$ axis.    As the time approaches the moment of impact, the radius and the vertical coordinate of the center of the circle decrease.  During the impact process, the back (right) side of the wave crest becomes close to a straight line, which has nearly a constant slope over time (about 15$^\circ$). At the moment of the impact, the circular arc shrinks  to a  point and the entire crest region takes on the shape of a straight line sloping down from the contact point toward the wave maker.  This final shape is the same in both plots as is the corresponding height of the contact point, $y\approx 0.11 \lambda_0$.

\begin{figure}[!htb]
\begin{center}
\begin{tabular}{c}
(a)  \\
\includegraphics[width=0.75\linewidth]{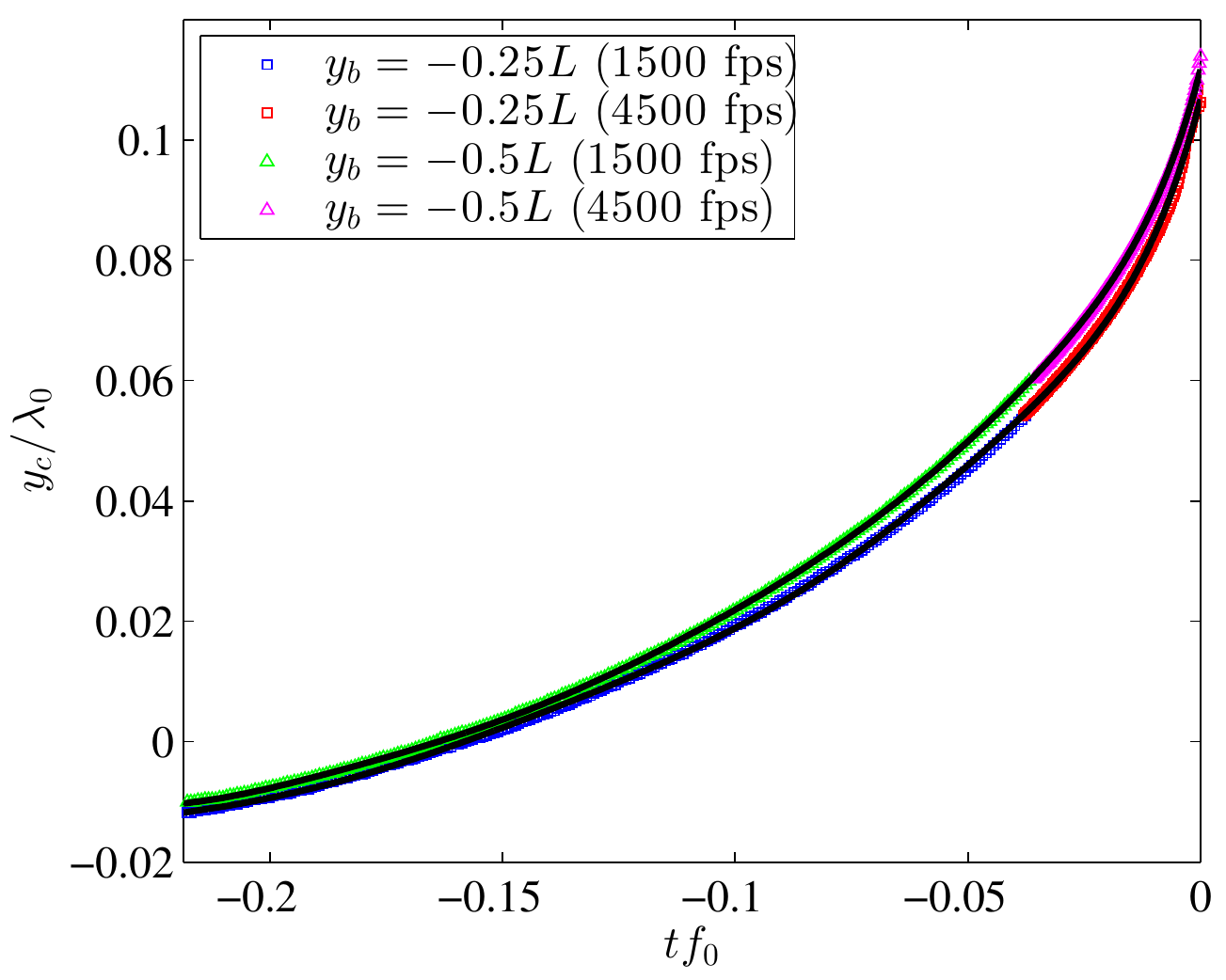} \\
(b) \\
 \includegraphics[width=0.75\linewidth]{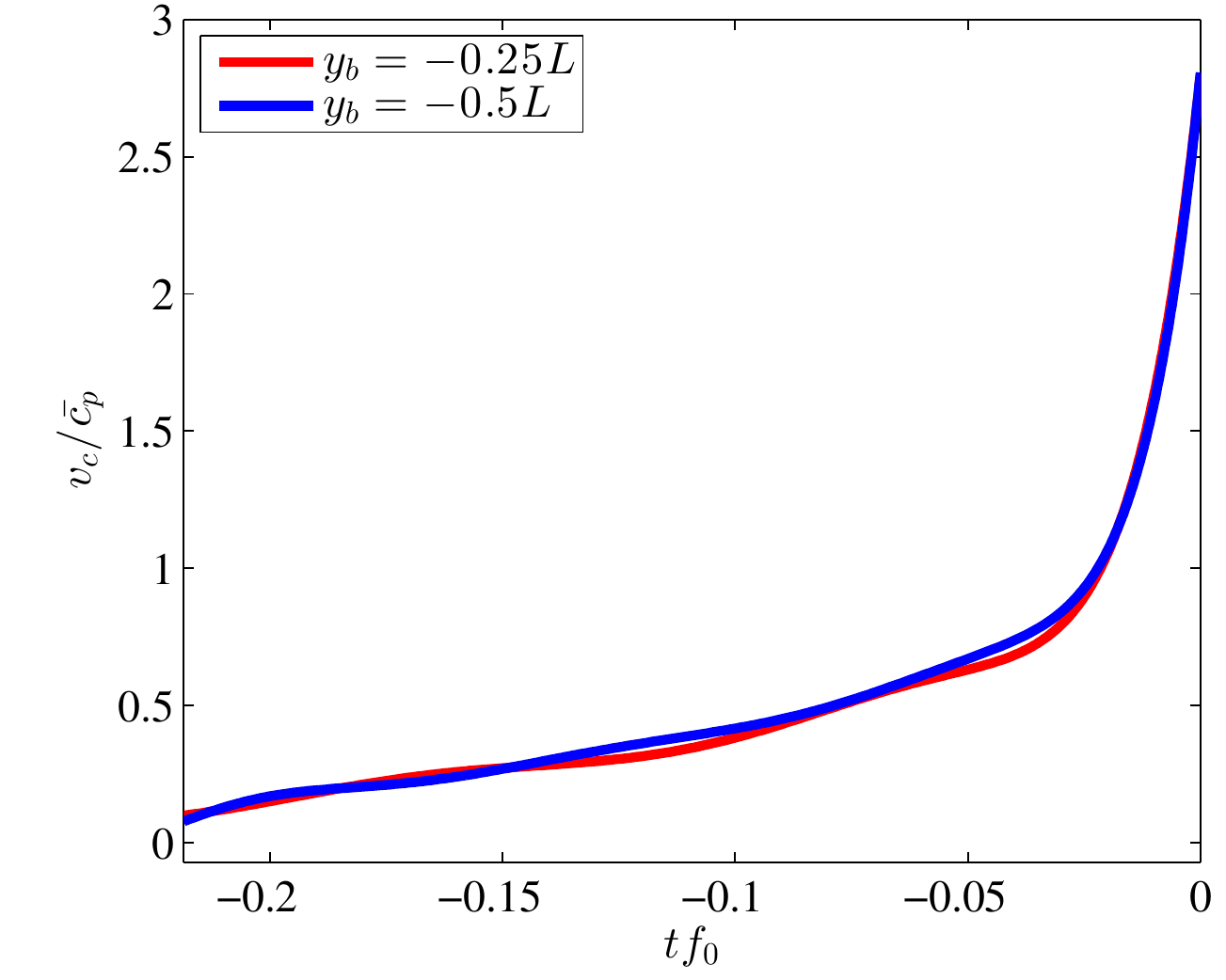} \\
(c) \\
\includegraphics[width=0.75\linewidth]{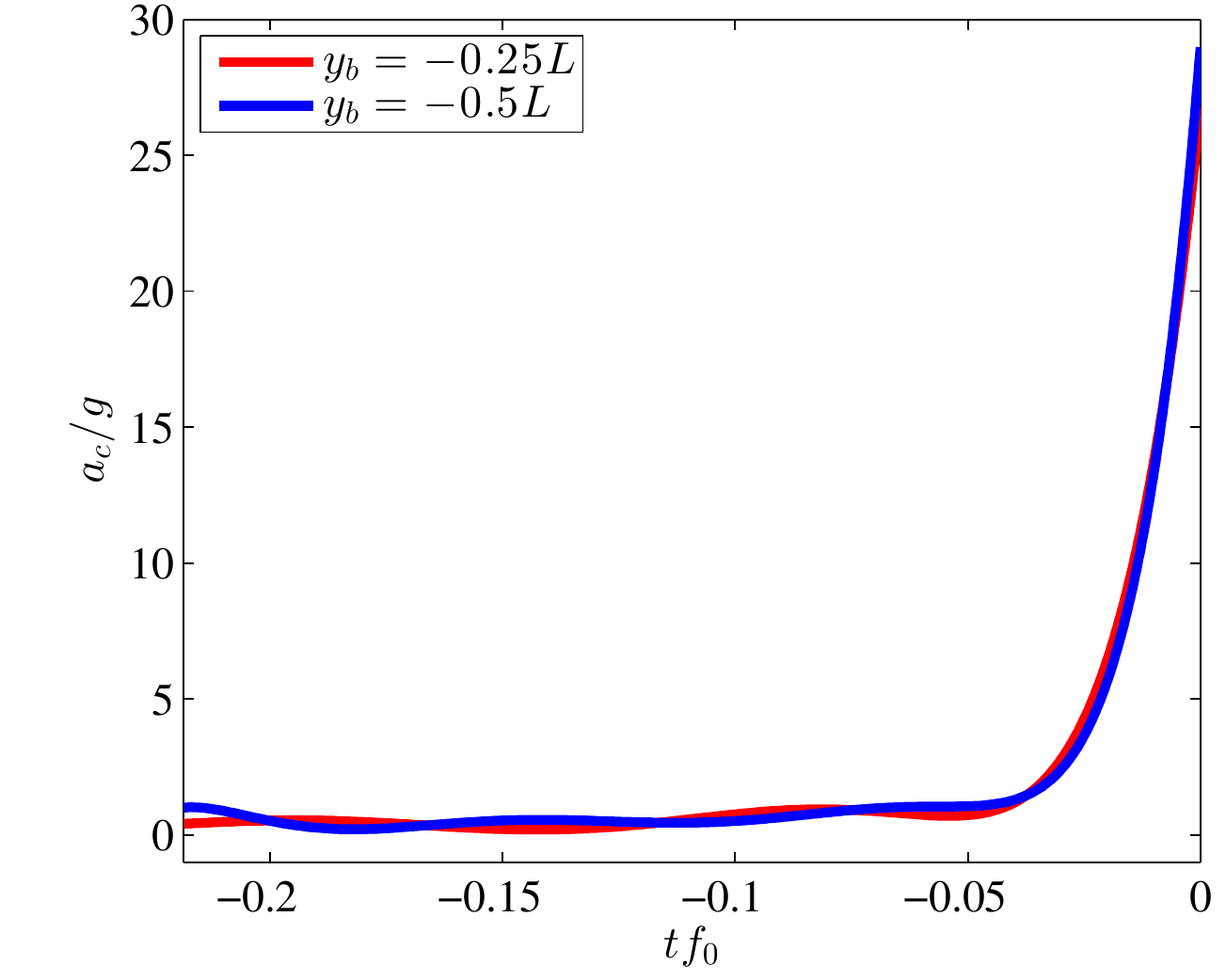}
\end{tabular}
\end{center}
\vspace{-0.2in} \caption{Motion of the contact point. The time is measured from the moment of impact. (a) Vertical position of the contact point, $y_c$. The blue and red markers represent data points for condition  $y_{b}$ = -0.25$L$ while. The green and magenta markers represent data points for condition $y_{b}$ = -0.5$L$. The black curves are the polynomial fits to the data points. For the same submergence, the data points processed from the two different movies with different frame rates are plotted with different color. (b) The velocity of the contact point. The data points are evaluated from $dy_{c}/dt$ of the polynomial fit of $y_{c}$. (c) The acceleration of the contact point. The data points are evaluated from $d^{2}y_{c}/dt^{2}$ of the polynomial fit of $y_{c}$. In both (b) and (c), blue data points represent the $y_{b}$ = -0.25$L$ condition and red data points represent the $y_{b}$ = -0.5$L$ condition.} \label{fig:contactpt}
\end{figure}

The the vertical position, $y_c$, the vertical velocity, $v_c=dy_c/dt$, and the vertical acceleration, $a_c=d^2y_c/dt^2$, of the water surface contact point on the front face of the cube are plotted in Figures \ref{fig:contactpt}(a), (b) and (c), respectively.   The horizontal axes are time given in wave periods ($tf_0$) with the moment of impact taken as $tf_0=0$.  As can be seen in Figure\ref{fig:contactpt}(a),  the contact point height  increases continuously with time  for both cube positions.  The continuous black lines in the plot are 9th order polynomials that were fitted to the data.  
The contact point velocities and accelerations plotted Figure~\ref{fig:contactpt}(b) and (c), respectively,  are obtained by differentiation of these polynomials.
The contact point velocity, $v_c$, in Figure~\ref{fig:contactpt}(b) is nondimensionlized by the average phase speed of the wave packet, $\overline{c}_p = 1.437$~m/s.  For $tf_{0}< 0.0345$ (30~ms), the contact point velocity increases with time in a nearly linear fashion.   At about 
$tf_{0}=0.0345$, this rate of change increases dramatically with the contact point velocity reaching about 2.75$\overline{c}_p$ in both cases, at the moment of impact.  
The contact point acceleration, $a_c$, in Figure~\ref{fig:contactpt}(c) is nondimensionlized by the gravitational acceleration, $g$. At about 
$tf_{0}=0.0345$, the contact point acceleration begins a rapid increase, reaching $28g$  at the moment of impact in both cases.
The high acceleration of the contact point indicates that a high pressure gradient exists in the bulk of the fluid just below the contact point. This high pressure gradient drives the high-speed motion of the contact point, which initiates the high-speed vertical jet after the moment of impact. 

\begin{figure*}[!htb]
\begin{center}
\begin{tabular}{cc}
(a) & (b) \\
\includegraphics[width=0.47\linewidth]{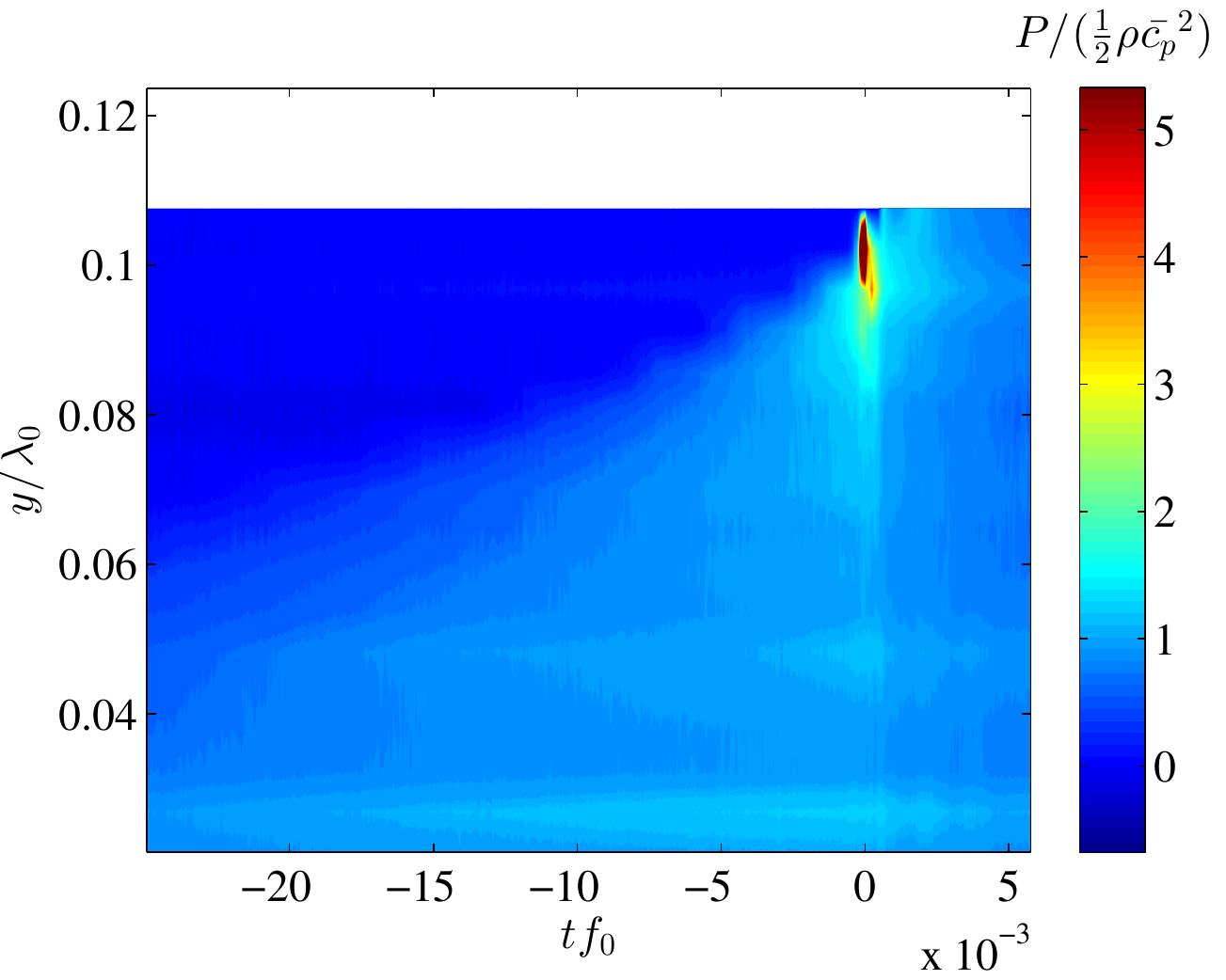} & \includegraphics[width=0.47\linewidth]{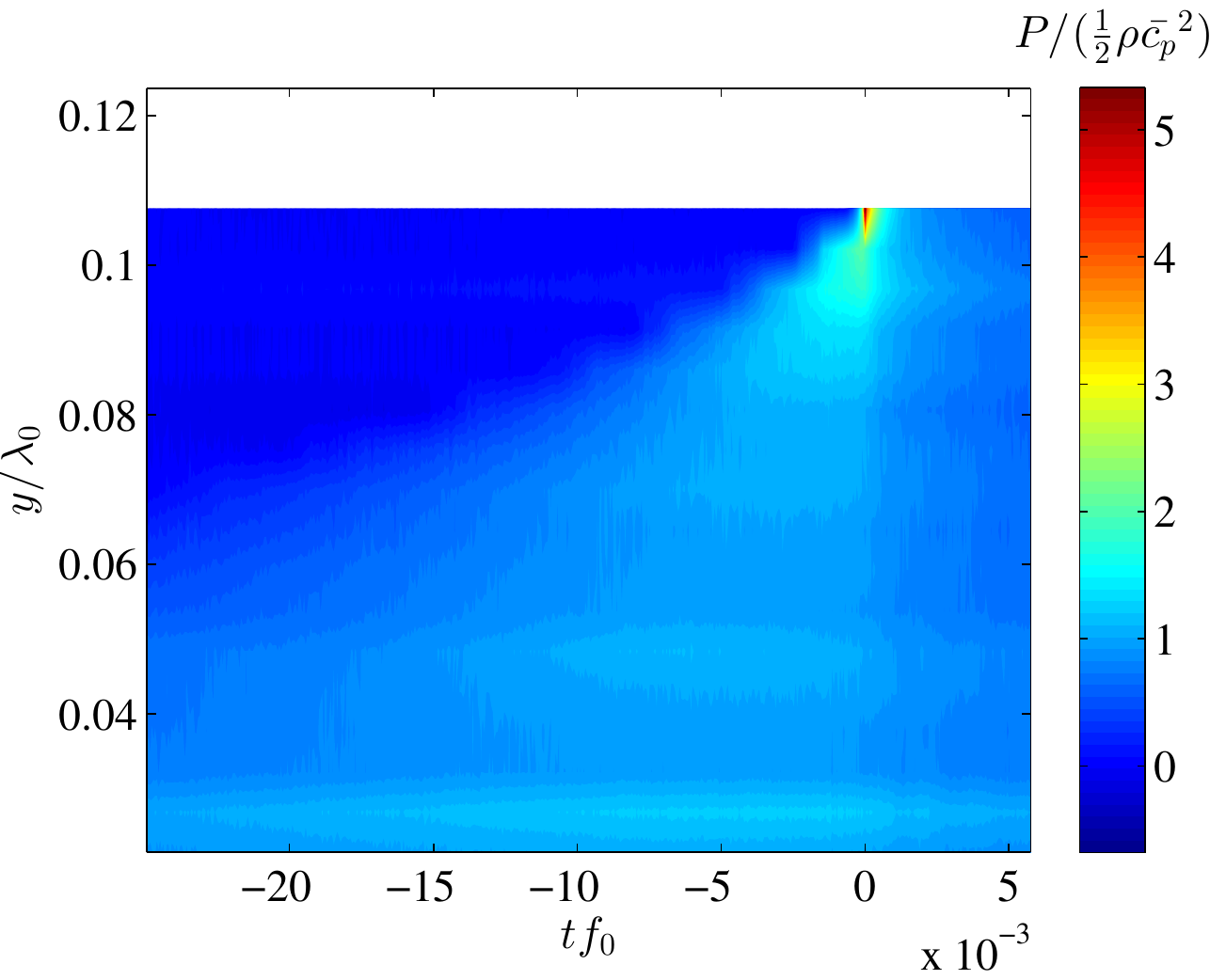} \\
(c) & (d) \\
\includegraphics[width=0.47\linewidth]{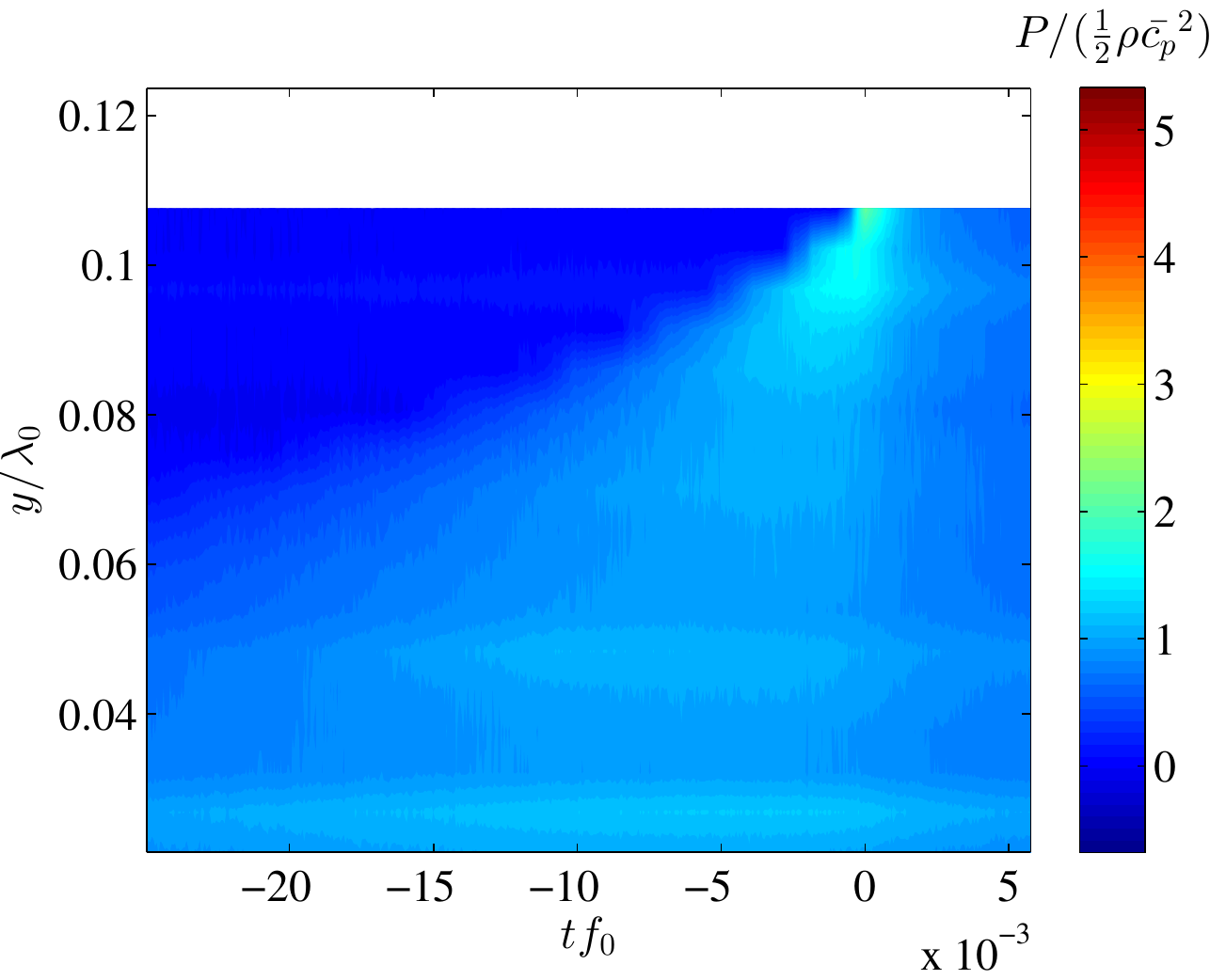} & \includegraphics[width=0.47\linewidth]{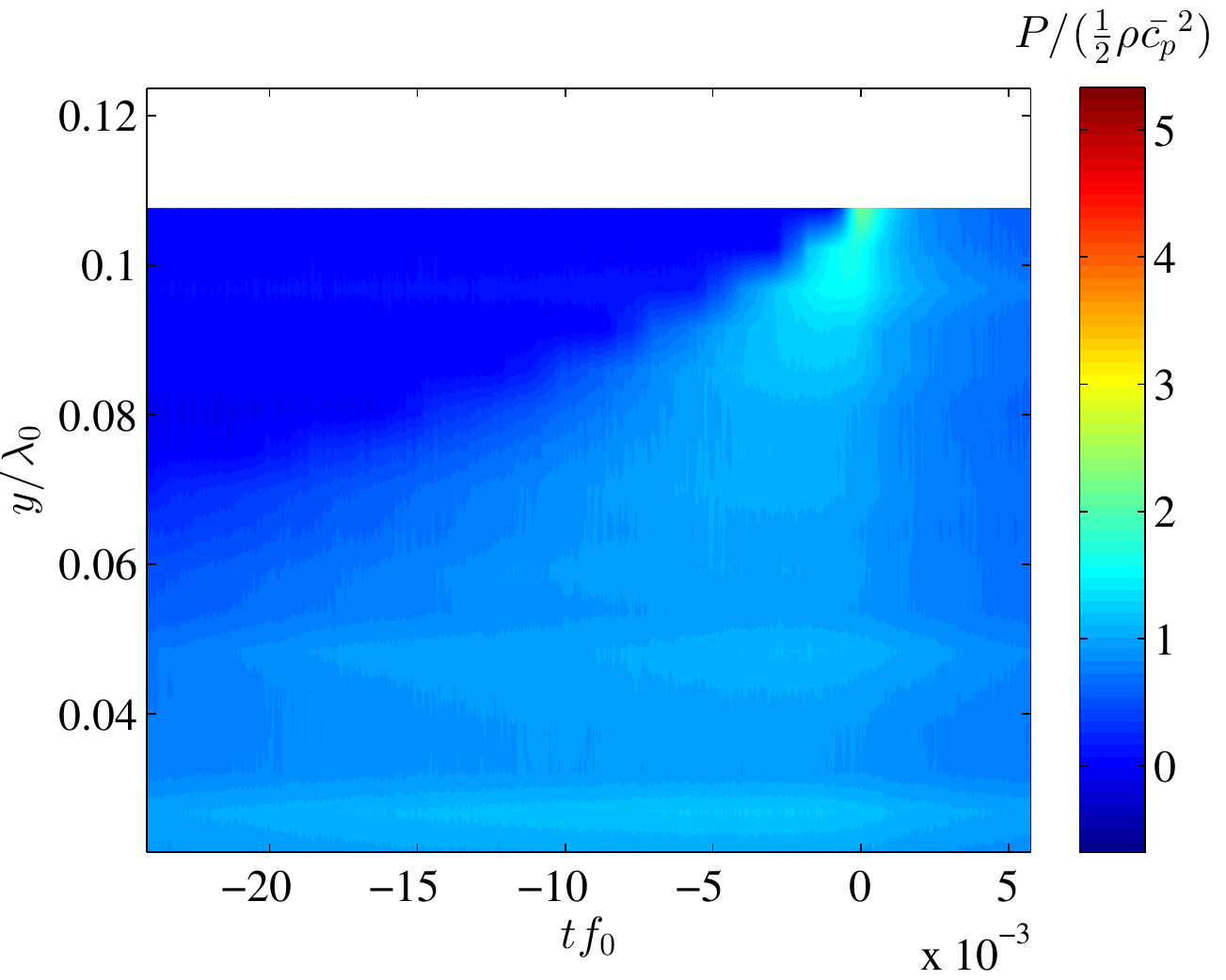} \\
\end{tabular}
\end{center}
\vspace*{-0.2in} \caption{Temporal evolution of the pressure distribution on the front face of the cube for four  experimental runs for the case $y_{b}$ = -0.5$L$. Subplots (a) to (d) correspond to experimental runs 1 to 4, respectively. The vertical axis is nondimensional position along the front face of the cube ($y/\lambda_0$) with $y/\lambda_0=0$ at the undisturbed water level.  The horizontal axis is dimensionless time ($tf_0$) with $tf_0=0$ taken as the moment of impact.  
The contour level represents the magnitude of the dimensionless pressure, $2P/\rho c_0^2$.} \label{fig:contour_sub15.24}
\end{figure*} 

\begin{figure*}[!htb]
\begin{center}
\begin{tabular}{cc}
(a) & (b) \\
\includegraphics[width=0.47\linewidth]{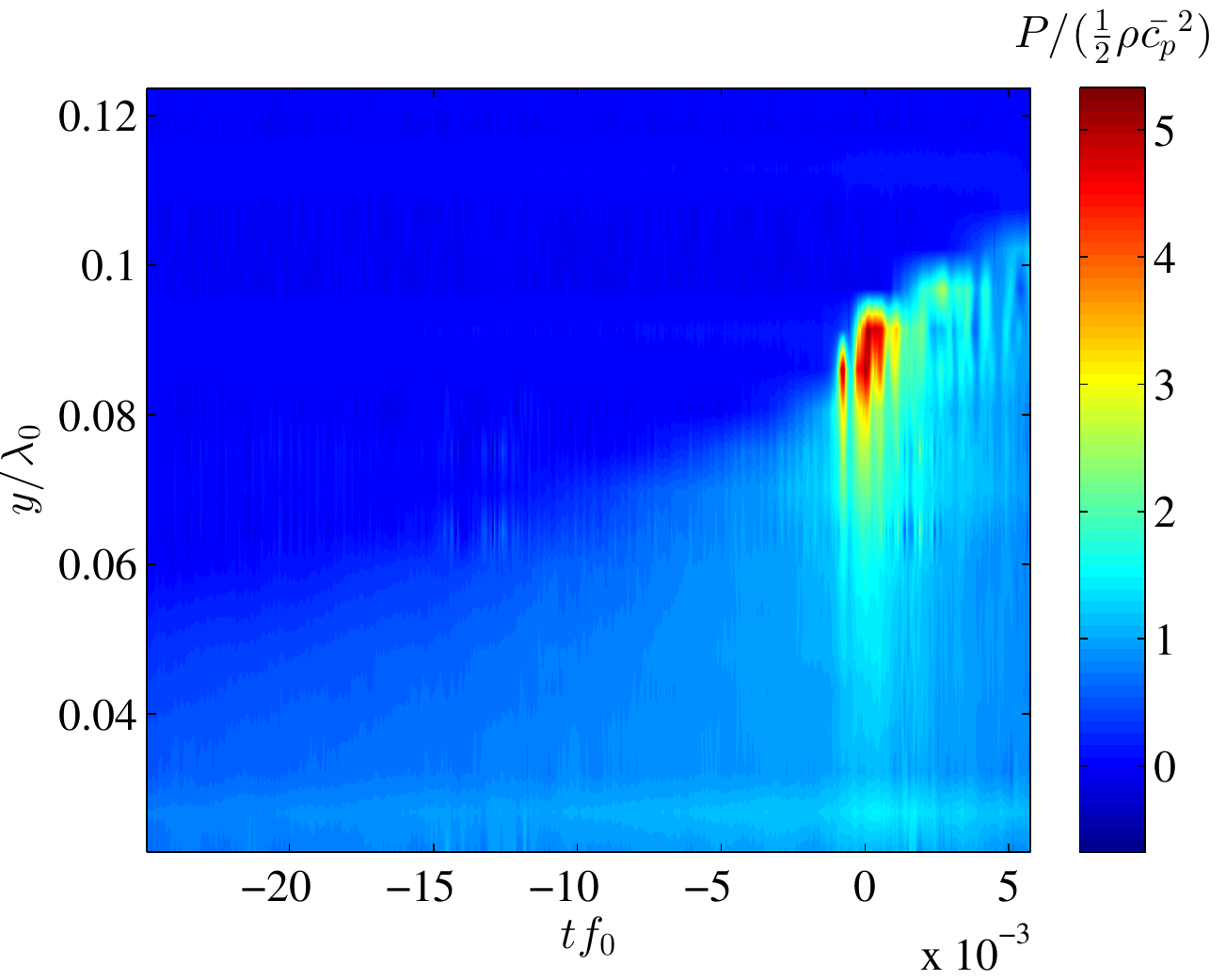} & \includegraphics[width=0.47\linewidth]{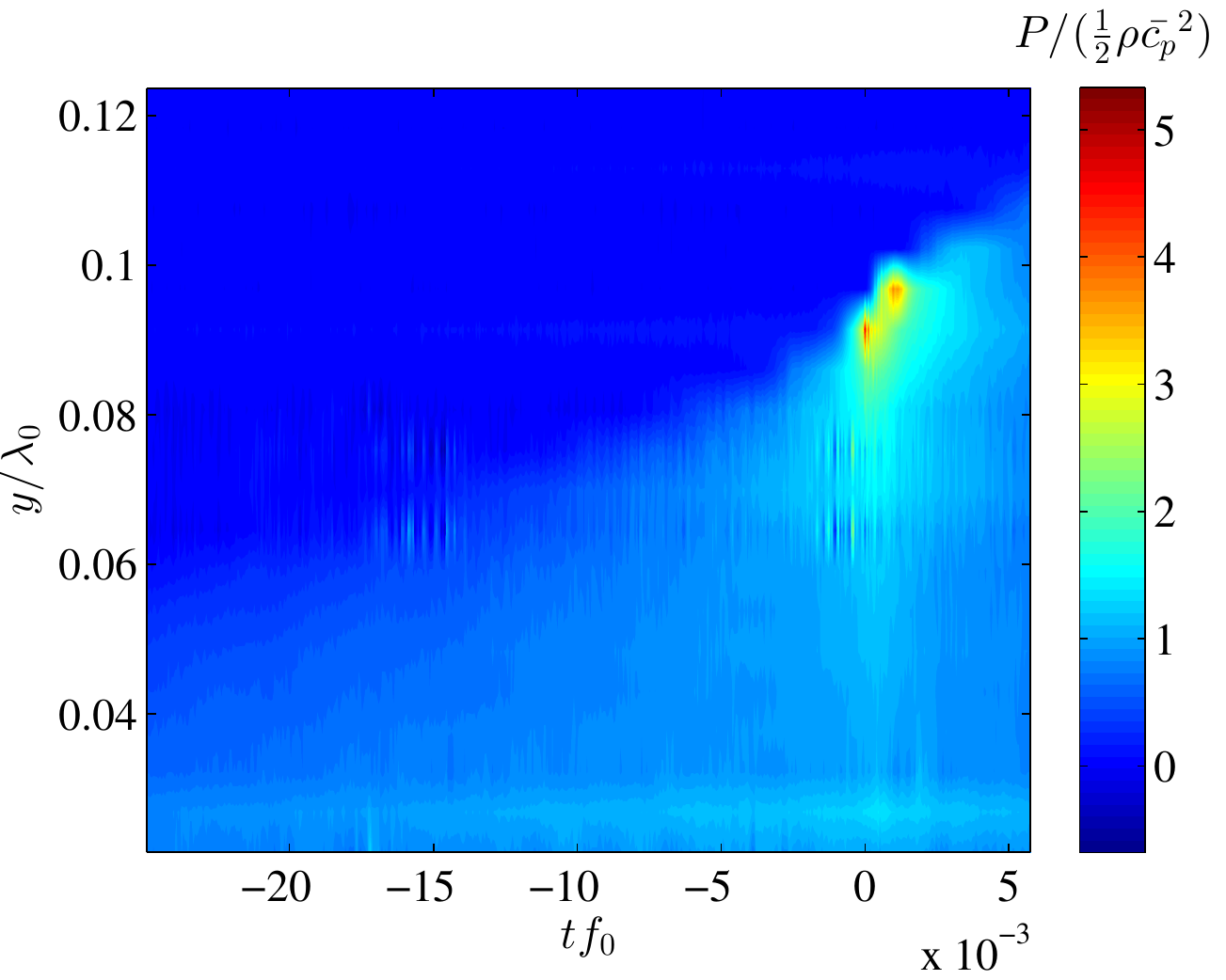} \\
(c) & (d) \\
\includegraphics[width=0.47\linewidth]{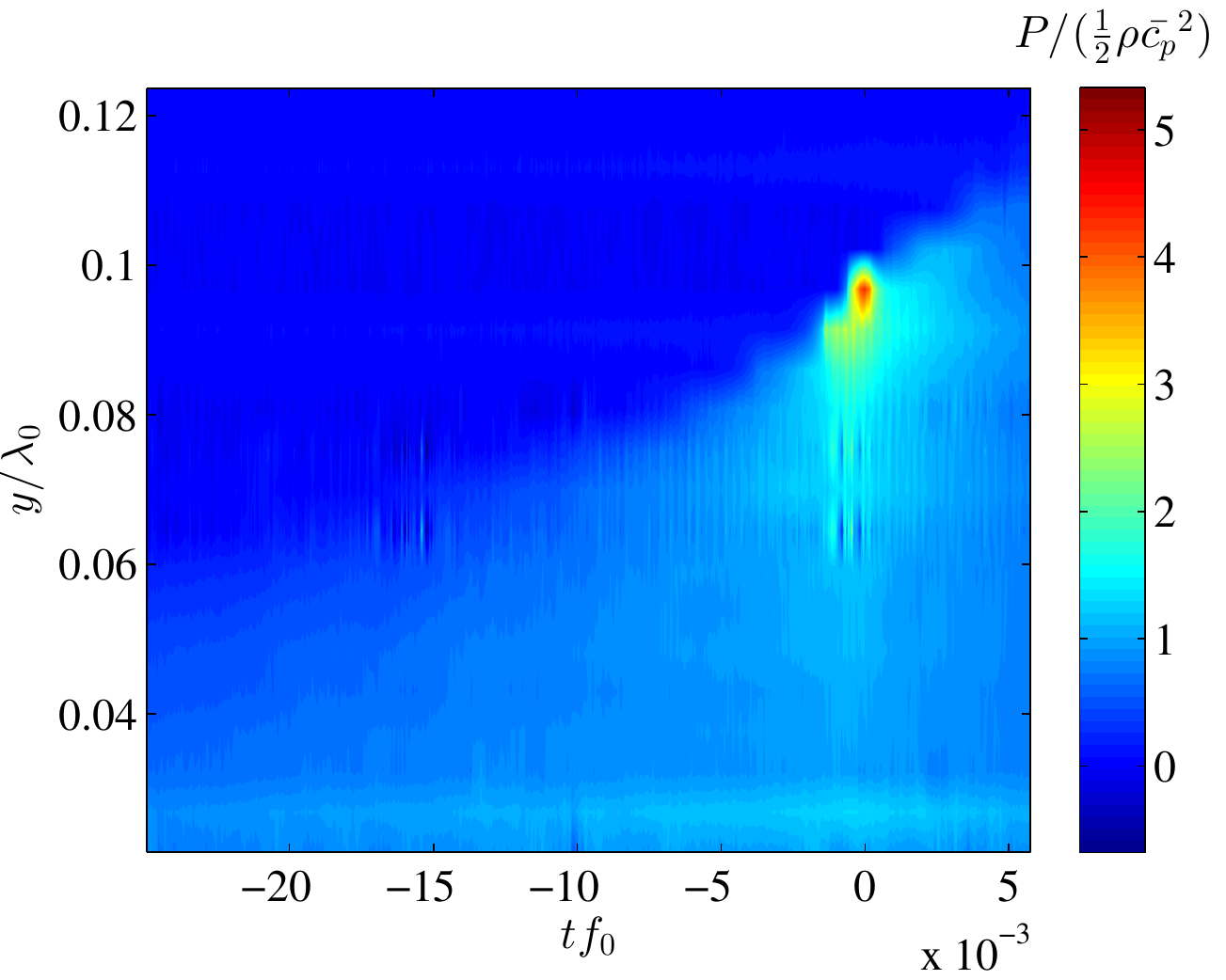} & \includegraphics[width=0.47\linewidth]{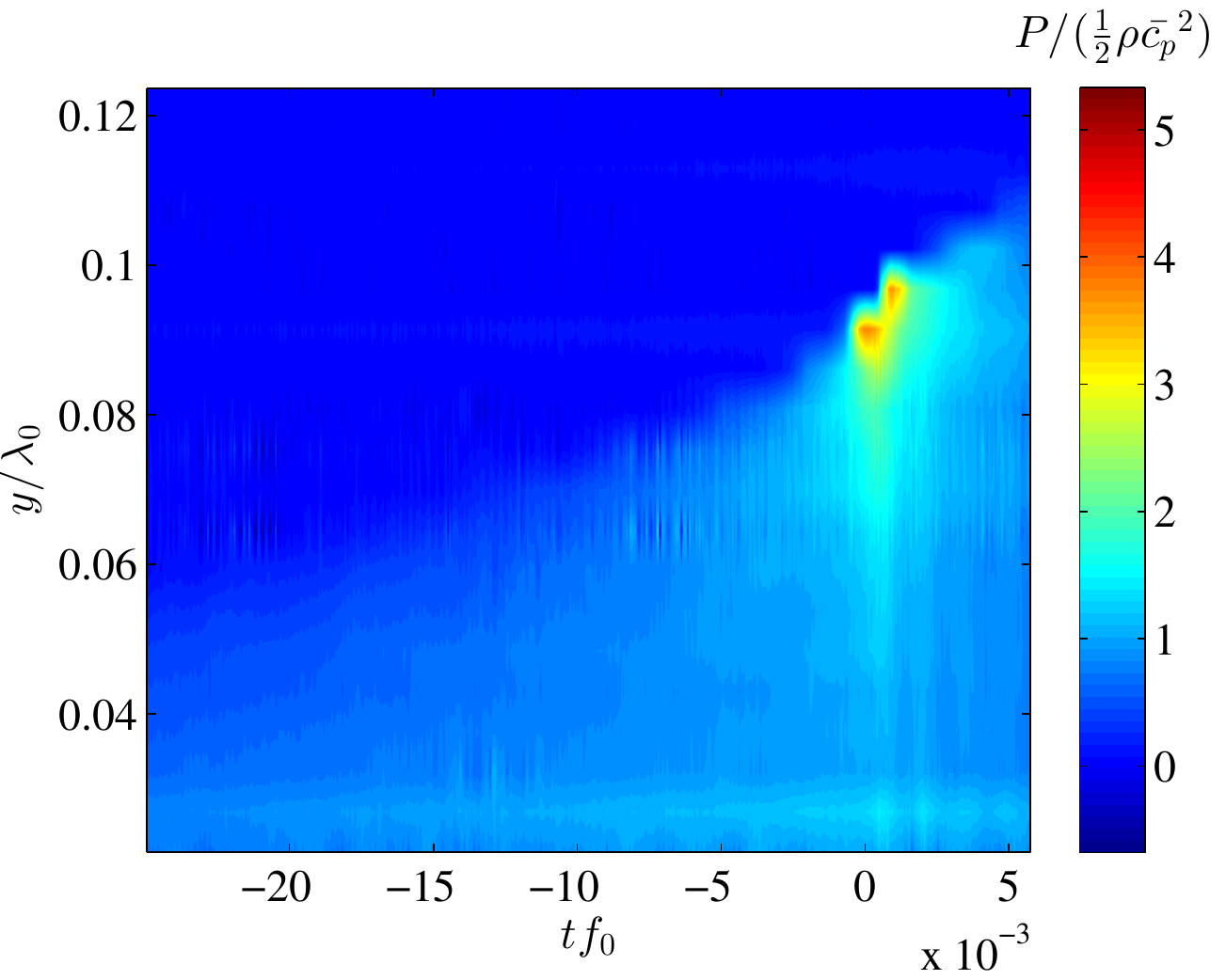} \\
\end{tabular}
\end{center}
\vspace*{-0.2in} \caption{Temporal evolution of the pressure distribution on the front face of the cube for four  experimental runs for the case $y_{b}$ = -0.25$L$. Subplots (a) to (d) correspond to experimental runs 1 to 4, respectively. The vertical axis is nondimensional position along the front face of the cube ($y/\lambda_0$) with $y/\lambda_0=0$ at the undisturbed water level.  The horizontal axis is dimensionless time ($tf_0$) with $tf_0=0$ taken as the moment of impact.  
The contour level represents the magnitude of the dimensionless pressure, $2P/\rho c_0^2$.} \label{fig:contour_sub7.62}
\end{figure*} 

\begin{figure*}[!htb]
\begin{center}
\begin{tabular}{cc}
(a) & (b) \\
\includegraphics[width=0.47\linewidth]{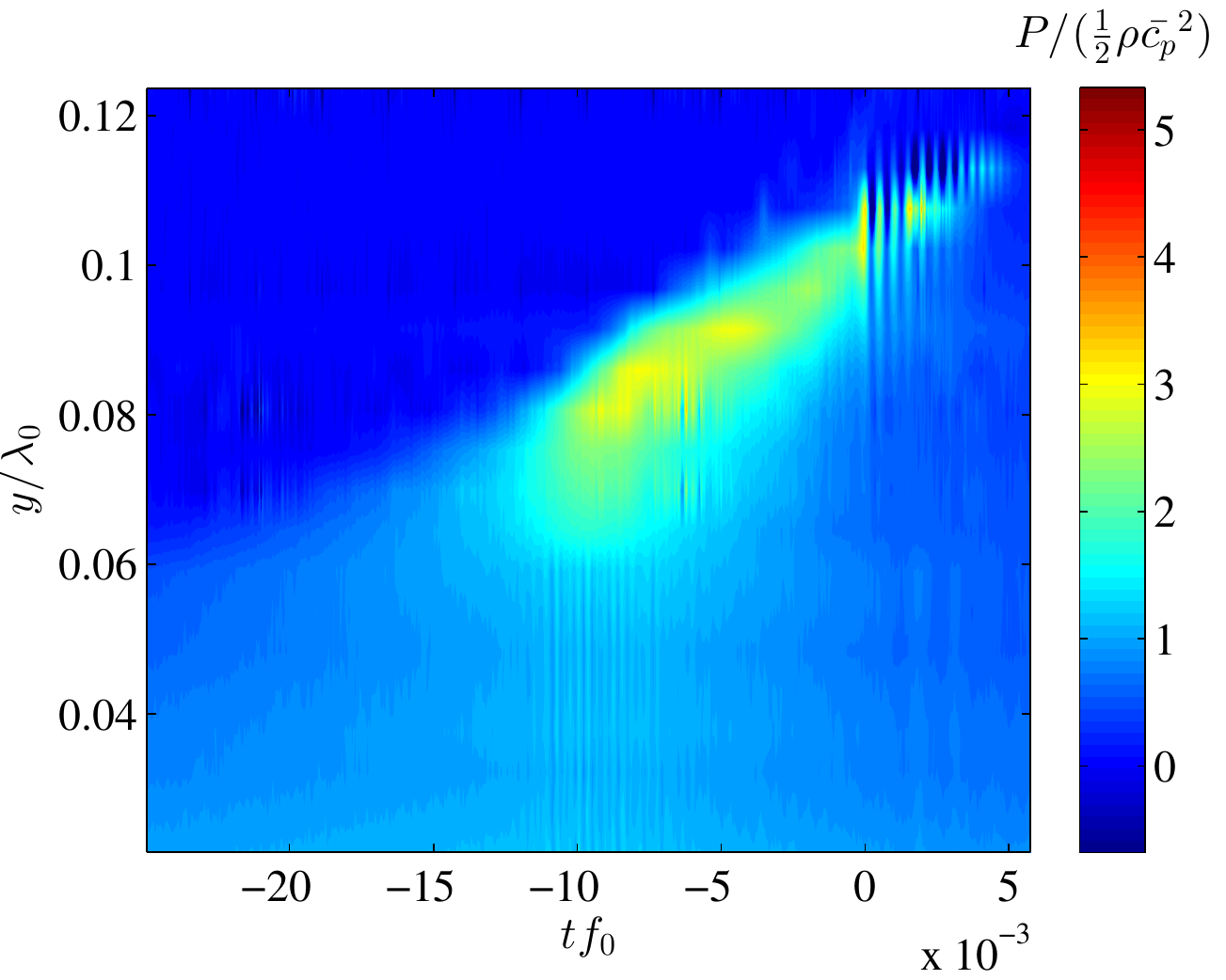} & \includegraphics[width=0.47\linewidth]{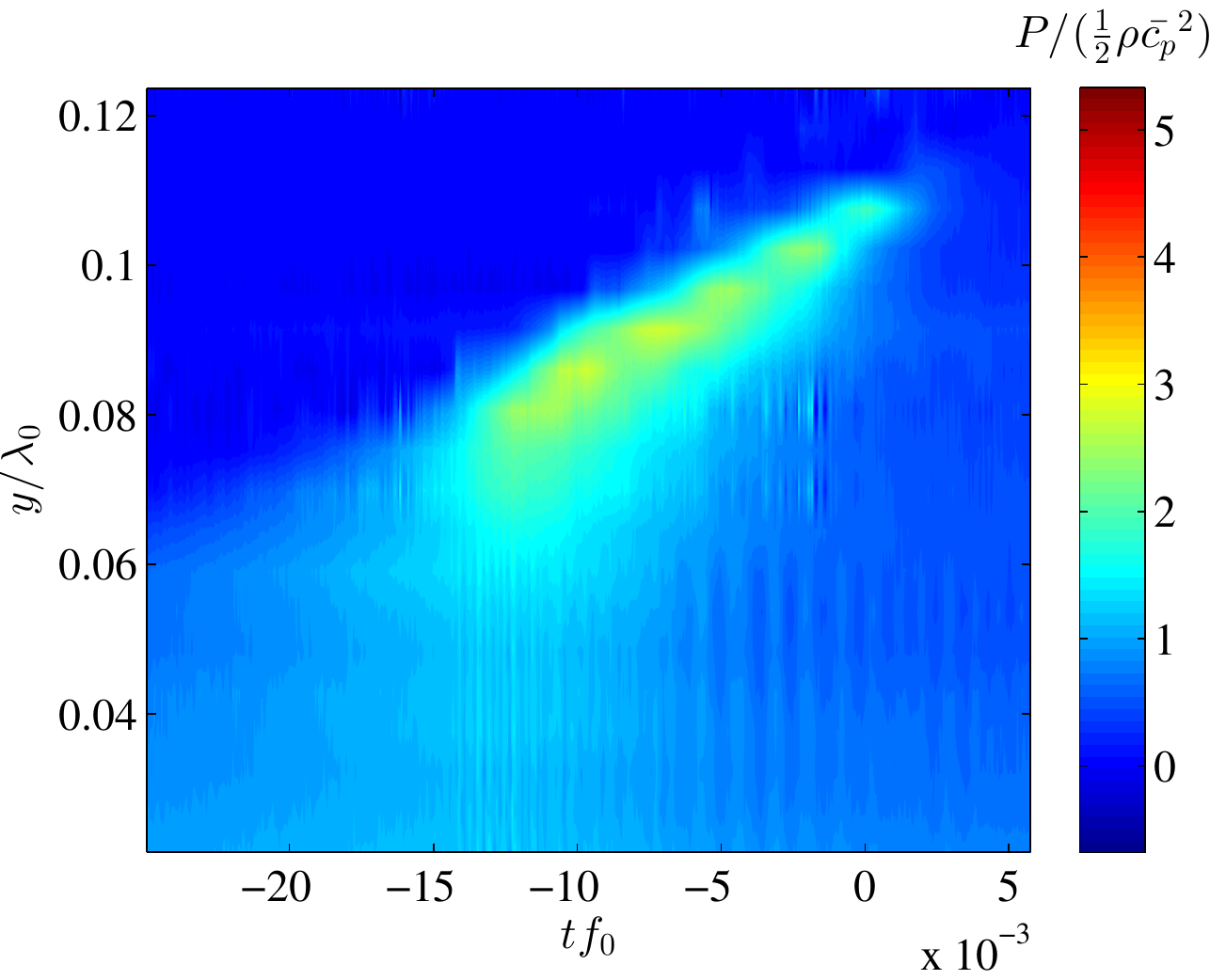} \\
(c) & (d) \\
\includegraphics[width=0.47\linewidth]{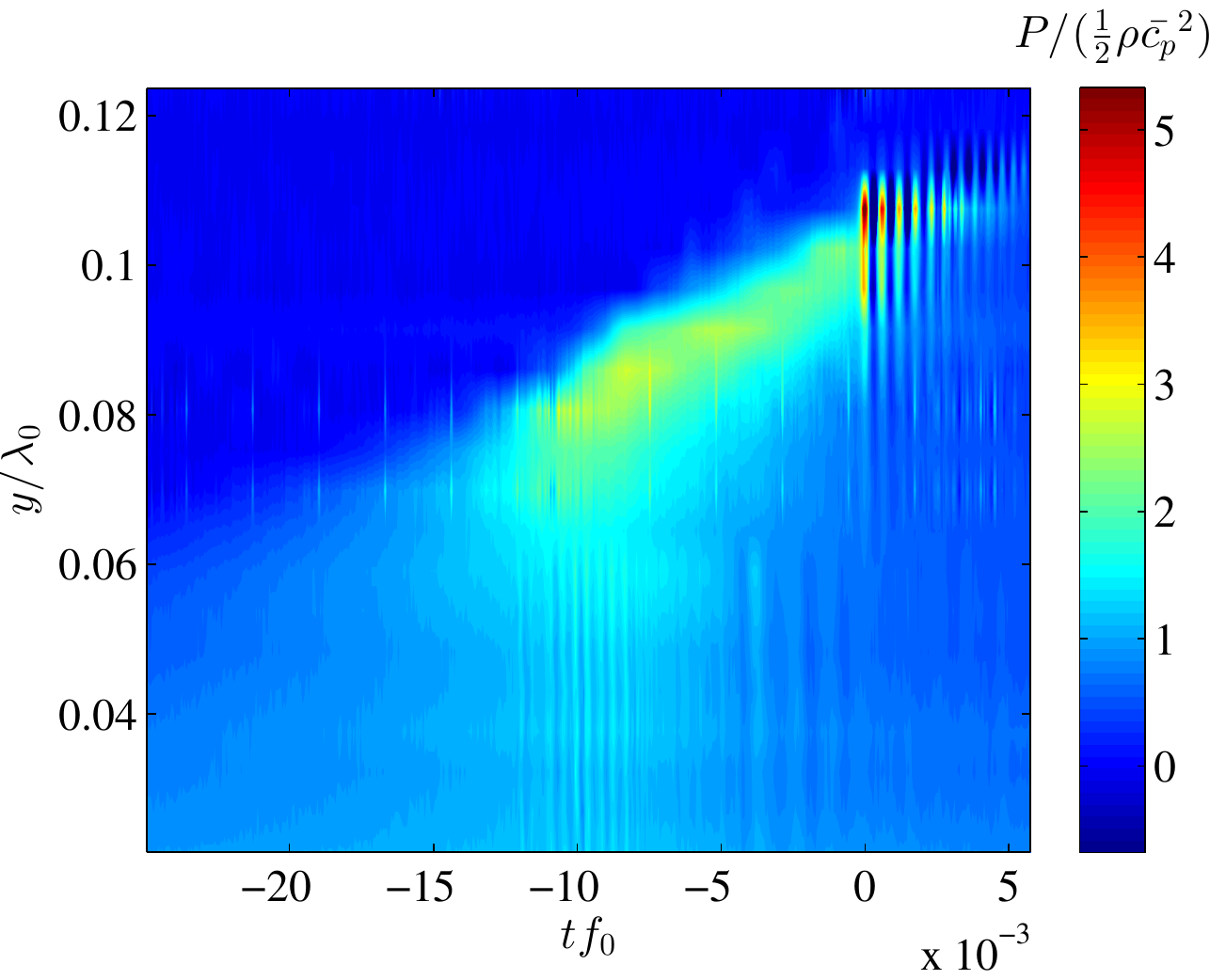} & \includegraphics[width=0.47\linewidth]{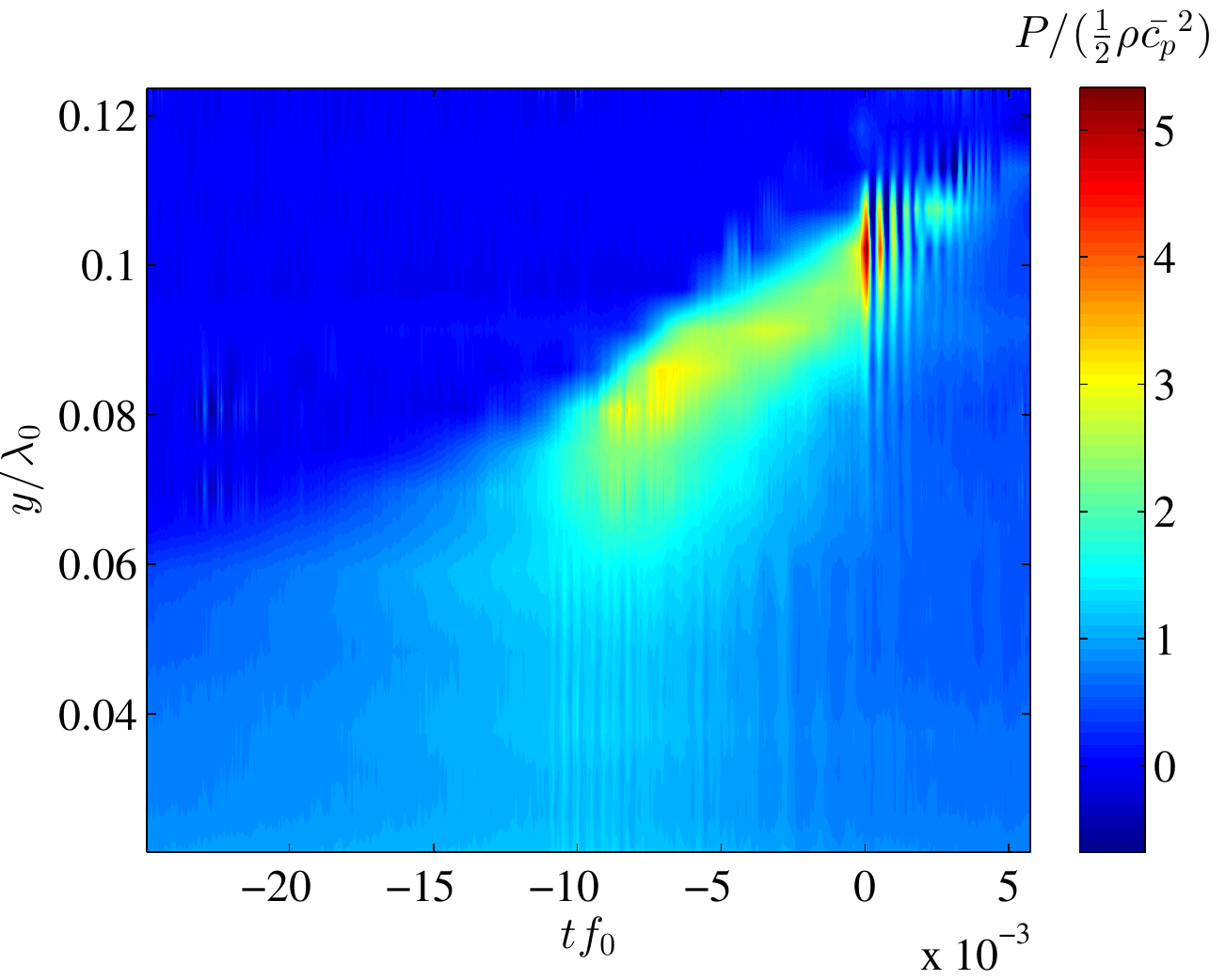} \\
\end{tabular}
\end{center}
\vspace*{-0.2in} \caption{Temporal evolution of the pressure distribution on the front face of the cube for four  experimental runs for the case $y_{b}$ = 0. Subplots (a) to (d) correspond to experimental runs 1 to 4, respectively. The vertical axis is nondimensional position along the front face of the cube ($y/\lambda_0$) with $y/\lambda_0=0$ at the undisturbed water level.  The horizontal axis is dimensionless time ($tf_0$) with $tf_0=0$ taken as the moment of impact.  
The contour level represents the magnitude of the dimensionless pressure, $2P/\rho c_0^2$.}  \label{fig:contour_sub0}
\end{figure*}

\begin{figure*}[!htb]
\begin{center}
\begin{tabular}{cc}
(a) & (b) \\
\includegraphics[width=0.45\linewidth]{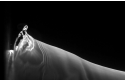} & \includegraphics[width=0.45\linewidth]{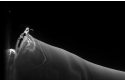} \\
(c) & (d) \\
\includegraphics[width=0.45\linewidth]{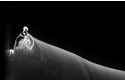} & \includegraphics[width=0.45\linewidth]{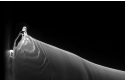} \\
\end{tabular}
\end{center}
\vspace*{-0.2in} \caption{Images of the  water surface for four experimental runs at condition $y_{b}$ = 0. The images were taken about 11 ms ($tf_{0}$ = 0.01265) before the moment of impact. The field of view of the images is 14.2~cm by 9.1~cm. Images (a) through (d) correspond to runs 1 through 4, respectively.} \label{fig:watersurface_repeated_sub0}
\end{figure*}

Figures  \ref{fig:contour_sub15.24}, \ref{fig:contour_sub7.62} and  \ref{fig:contour_sub0} show the temporal evolution of pressure distribution on the front face of the cube for the $y_{b}$ = -0.5$L$, $y_{b}$ = -0.25$L$ and $y_{b}$ = 0 cases, respectively. In these plots, the horizontal axis is dimensionless time, $tf_0$, and the vertical axis is dimensionless position along the front face of the cube, $y/\lambda_0$, with $y/\lambda_0=0$ at the mean water level, i.e.\ a different position on the cube for each figure.  The dimensionless pressure, $2P/(\rho c_p^2)$, where $\rho$ is the density of water, is given as color contours according to the color bar to the right of each plot.  Each figure includes separate plots from four experimental runs.

In all the pressure contour plots, the dark-blue region in the upper left corner corresponds to the sensors that are above the instantaneous water surface and thus measure atmospheric pressure, $P=0$.  
Starting at the left side of the plots, the lower light blue region corresponds to the part of the cube face that is below the water level.  Initially, the pressures are slightly higher than atmospheric.  As time increases, the top of this region increases in height.  The highest pressures due to the wave impact occur close to the instantaneous water level and the timing, magnitude and vertical extent of this high-pressure region varies from one experimental condition to another.  For the $y_b=-0.5L$ case, see Figures ~\ref{fig:contour_sub15.24}, the the high-pressure region is confined to a small region with a width of only about $0.01/f_0 = 9$~ms.  The highest pressures occur at or near the top of the measurement region; however, since the center of the top sensor is 25.4~mm from the top edge of the front face of the cube, it is not possible to give an accurate measure of the vertical extend of the high-pressure region.

For the $y_{b} = -0.25 L$ case, see Figure~\ref{fig:contour_sub7.62}, the shape of the high-pressure region is broader, about $0.025/f_0$, than that for $y_b=-0.5L$.   The top of the cube is 22.86~cm $ = 0.194\lambda_0$ above the mean water level, which is well above the location of the high-pressure region.  The data in the figure indicates that the vertical extent of the region is about 0.01$\lambda_0 = 1.2$~cm.  A double pressure peak is observed for this  condition.  The two peaks are probably due to the impact of the small jet mentioned above and the fluid motion in the  flip-through impact.  According to Figure \ref{fig:evolutions_sub7.62}, the small jet hits cube face as the contact point reaches this impact location.    The small separation of this two events in time and space is thought to correspond to the time-space separation of  the two pressure peaks in Figure \ref{fig:contour_sub7.62}. By comparing the maximum pressure at the impact location from different experimental runs, one can find that the magnitude of the peak pressure is not exactly repeatable, even though the  pressure distributions earlier in time are quite repeatable. The maximum pressure at the impact location appears to be strongly affected by the small  features in the water surface shape, which seem to originate from the water surface contact point  earlier in the impact process.  
These small features are amplified when the nearly circular surface profile shrinks towards a point. 
Even though these features have small dimension, the surface velocities are  quite large, which can result in high impact pressures when they hit the wall.

For the $y_b=0$ case, see Figure~\ref{fig:contour_sub0}, the shape of the high-pressure region is quite different than in the other two cases.  Moderately high pressures begin early, at about -0.1/$f_0$, and extend over a wider vertical range.  This vertical range narrows as time proceeds to the moment of impact.  In subplots (a), (c) and (d), the moment of impact is the beginning of a high-pressure oscillation that extends over a depth of about $0.02\lambda_0$ and has a frequency of about 2000~Hz.  This is likely due to oscillations of an entrapped air bubble as  was speculated to occur from examination of the high-speed movies.  This pressure oscillation does not occur for the run shown in sub-plot (b).  A reason for this difference in behavior was explored unsuccessfully by examination of the corresponding high-speed movies.  As can be seen in the single images taken just before the moment of impact and shown in Figure~\ref{fig:watersurface_repeated_sub0}, from the point of view of this single camera, the impacts look nearly identical.

Cooker and Peregrine (1995)  proposed a pressure-impulse theory for modeling the impact pressure and velocity field. In the present paper, the pressure impulse is defined as
\begin{equation}
I=\int^{t_{2}}_{t_{1}}P\left(t\right)dt
\end{equation}
where $t_{1}$ is the moment of time when the pressure starts to rise and $t_{2}$ is the moment of time when the pressure reaches the maximum. The pressure rise time is defined as
\begin{equation}
t_{r}=t_{2}-t_{1}
\end{equation}

\begin{figure*}[!htb]
\begin{center}
\begin{tabular}{cc}
(a) & (b) \\
\includegraphics[width=0.45\linewidth]{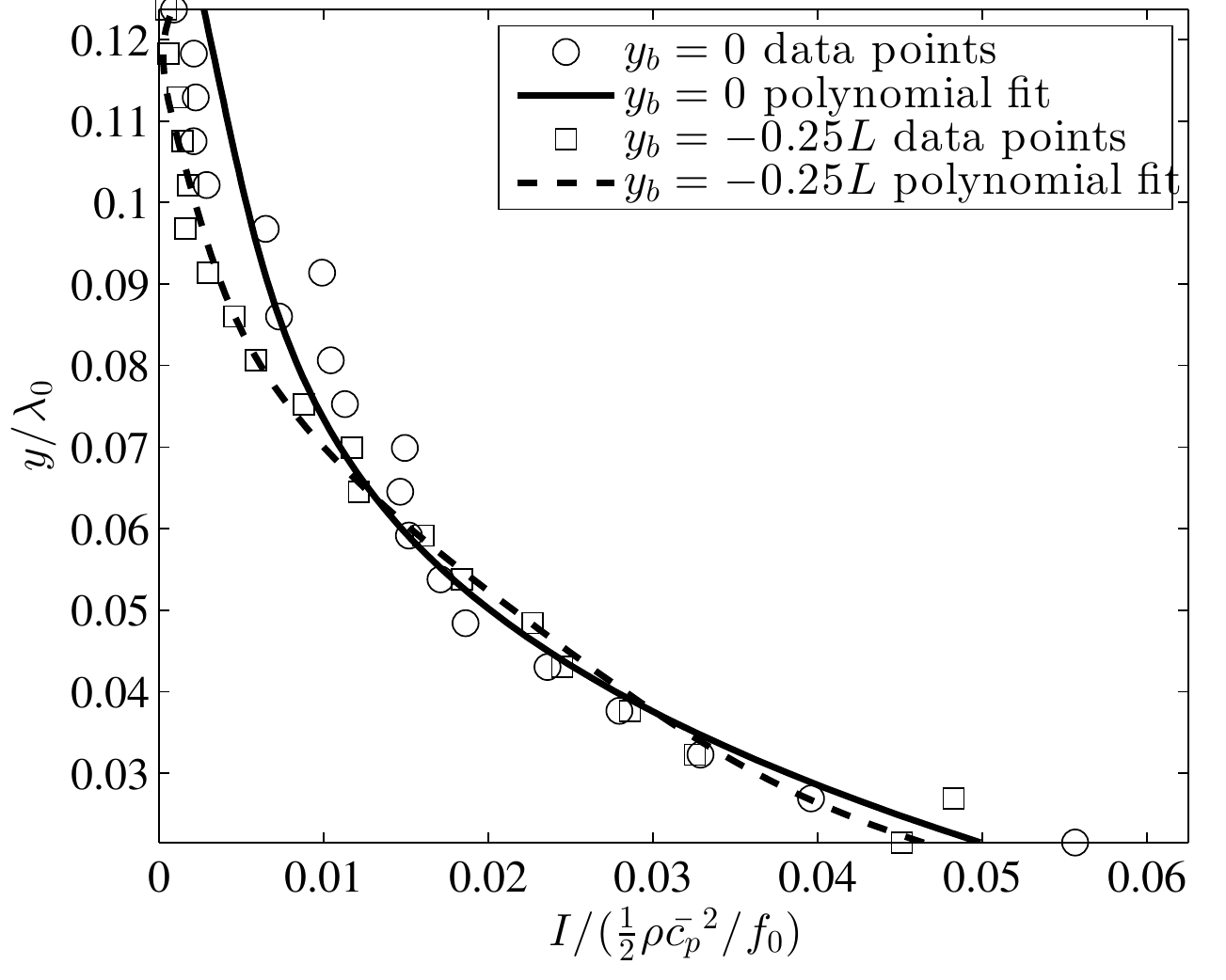} & \includegraphics[width=0.45\linewidth]{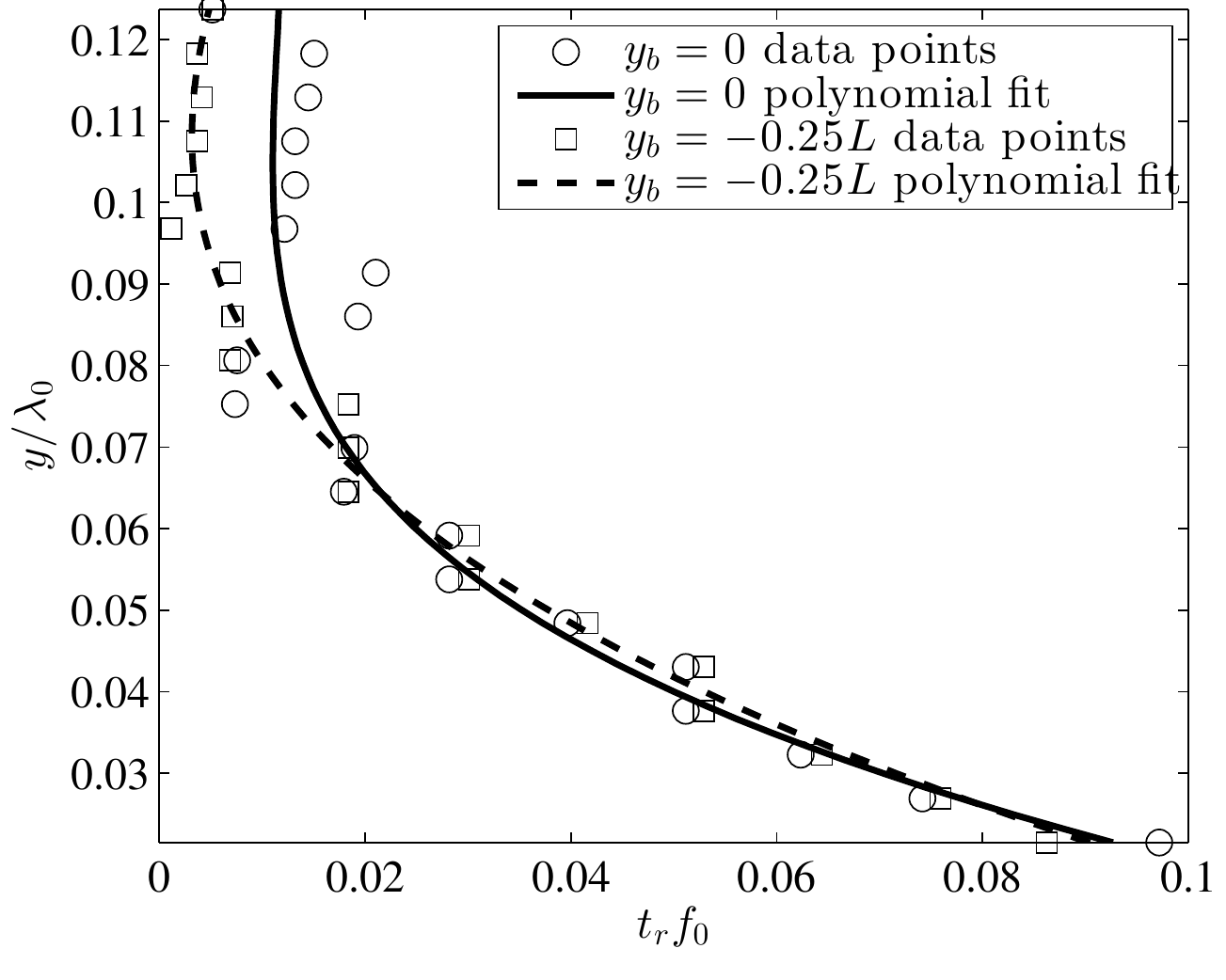} \\
\end{tabular}
\end{center}
\vspace*{-0.2in} \caption{Pressure impulse and rise time versus vertical position along the front face of the cube for the cases with $y_b=-0.25L$ and $y_b=0$.  (a) The pressure impulse versus height.   (b) The pressure rise time versus height. In both plots, the square markers and circle markers represent the data points from condition $y_{b}$ = -0.25$L$ and condition $y_{b}$ = 0, respectively. The dashed line and solid line represent the polynomial fit to the data points from condition $y_{b}$ = -0.25$L$ and condition $y_{b}$ = 0, respectively.} \label{fig:impulse_risetime}
\end{figure*} 

The vertical distributions of pressure impulse and rise time are shown in Figure \ref{fig:impulse_risetime}(a) and (b), respectively, for the $y_b = -0.5L$ and $-0.25L$ cases. For both cases, the rise time increases dramatically from higher vertical positions to lower vertical positions.  The pressure impulse follows the same trend, indicating that the  distribution of rise time dominates the distribution of the pressure peak value, which as seen in Figures~\ref{fig:contour_sub15.24} and \ref{fig:contour_sub7.62},  increases with increasing vertical position.  The rise times for the two cube positions have very similar values for lower positions along the cube face. The rise time for the $y_{b}$ = 0 condition is longer than that for the  $y_{b}$ = -0.25$L$ condition for positions around the impact zone. This may be caused by the cushion effect of the air pocket which is probably trapped in the $y_{b}$ = 0 condition.   In any case, the impulse and rise time near the impact zone are a bit higher in the $y_b=0$ case.

\section{CONCLUSIONS}

The impact of a plunging breaker on a rigidly mounted partially submerged cube (dimension $L= 30.48$~cm) was studied experimentally.  Measurements were performed at three experimental conditions, all with the same wave maker motion creating a nominal breaker wavelength of $\lambda_0= 1.18$~m.  The cube was placed at one streamwise position, with its front face located $5.44\lambda_0$ horizontally from the wave maker, and three vertical positions,  $y_b = -0.5L$, $-0.25L$  and $0$, where $y_b$ is the vertical position of the bottom of the cube with $y_b$ positive upward and $y_b=0$ at the undisturbed water level. Simultaneous measurements of the pressure distribution on the front face of the cube and the adjacent water surface profile in the streamwise vertical center plane of the cube were performed.

It was found that the vertical position of the cube has a dramatic effect on the wave impact and the resulting surface pressure distribution.  At $y_b=-0.5L$, the water surface between the crest and the water surface contact point on the cube forms a nearly circular arc with upward curvature.  The arc shrinks towards a point as the crest approaches the cube and the time when this arc disappears is called the moment of impact.  As the arc shrinks, the contact point moves upwards with increasing velocity and acceleration, reaching accelerations as high as 30$g$ shortly before the moment of impact. Subsequent to the moment of impact, a high-speed vertical jet is formed.  This free surface behavior is known as flip-through. The pressure on the wet part of the cube surface  includes a small region of high pressure near the free surface for a short time just before the moment of impact.  It is thought that the corresponding high subsurface pressure gradient is responsible for the very high acceleration of the contact point.  

For $y_b=-0.25L$, the free surface behavior is similar to the case with  $y_b=-0.5L$. However, in addition to the shrinking arc behavior, a small jet is formed on the front side of the crest just before the moment of impact. Due to its speed, this small jet is expected to carry high horizontal momentum. In this case, the pressure distribution on the cube front face includes a double-peak.  It is thought that the first peak is due to the impact of the jet, while the second is due to the flip-through phenomenon found also in the $y_b=-0.5L$ case. The surface pressure rise time and surface pressure impulse increase with decreasing vertical position  on the cube face.

For $y_b=0$, the crest starts to bend forward and forms a curved water sheet before the impact. This curved water sheet moves horizontally towards the front face of the cube as the contact point moves upward.  The moment of impact for this condition is the time when the tip of the water sheet meets the upward moving contact point.   This process seems to entrap a packet of air between the water moving up the cube face and the water sheet from the wave crest.  A surface pressure oscillation with a frequency of about 2,000~Hz is observed under this condition. The pressure oscillation is believed to be related to the coupled effect of the entrained air packet and the water flow. The pressure rise time and the impulse around the impact area for this condition are larger than those for  $y_b=-0.25L$.

\section{ACKNOWLEDGEMENTS}
The support of the Office of Naval Research under grant N000141410305, contract monitor Dr.\ R. Joslin, is gratefully acknowledged.


\end{document}